\documentclass[twocolumn,english,aps,prb,showpacs,superscriptaddress,floatfix,longbibliography]{revtex4-1}
\usepackage[latin9]{inputenc}
\usepackage{bm}
\usepackage{amsmath}
\usepackage{amssymb}
\usepackage{graphicx}
\usepackage{esint}
\usepackage{babel}
\begin{document}

\title{Conserving approximations in cavity quantum electrodynamics: Implications
for density functional theory of electron-photon systems }

\author{I. V. Tokatly}

\email{ilya.tokatly@ehu.es}

\selectlanguage{english}%

\affiliation{Departamento de F\'isica de Materiales,
Universidad del Pa\'is Vasco UPV/EHU, Av. Tolosa 72, E-20018 San Sebasti\'an,
Spain }

\affiliation{IKERBASQUE, Basque Foundation for Science, E-48011 Bilbao, Spain}

\affiliation{Donostia International Physics Center (DIPC), Manuel de Lardizabal
5, E-20018 San Sebasti\'an, Spain}
\begin{abstract}
By analyzing the many-body problem for non-relativistic electrons
strongly coupled to photon modes of a microcavity I derive the exact
momentum/force balance equation for cavity quantum electrodynamics.
Implications of this equation for the electron self-energy and the
exchange-correlation potential of quantum electrodynamic time-dependent
density functional (QED-TDDFT) are discussed. In particular I generalize
the concept of $\Phi$-derivability to construct approximations which
ensure the correct momentum balance. It is shown that a recently proposed
optimized effective potential approximation for QED-TDDFT is conserving
and its possible improvements are discussed.
\end{abstract}
\maketitle

\section{Introduction}

In most typical situations in condensed matter and chemical physics
electromagnetic fields interacting with the matter can be treated
classically. In this standard approach the components of the electromagnetic
4-potential enter quantum dynamics of charged particles as external
(possibly self-consistent) classical parameters by producing classical
forces which drive the system out of equilibrium and control its dynamics.
However an impressive progress in the fields of cavity and circuit
quantum electrodynamics (QED) has opened a possibility to study phenomena
in which the quantum nature of electromagnetic fields become essential
and a strong coupling between electrons and confined photons play
a key role. Historically, a strong coupling to quantum electromagnetic
cavity modes was first realized for electrons in Rydberg atoms in
the cavity-QED \citep{RaiBruHar2001,MabDoh2002,Walter2006}. Further
progress was related to the development of the circuit-QED where the
regime of strong electron-photon coupling is achieved for mesoscopic
systems, such as quantum dots or superconducting qubits embedded into
microwave transmission line resonators \citep{Wallraff2004,Blais2004,Frey2012,Delbecq2011,Petersson2012,Liu2014}.
Recently the realm of the cavity/circuit-QED has been extended to
more complicated and reach electronic systems, such as organic molecules
in an emerging field of ``chemistry in cavity'' or a ``polaritonic
chemistry'' \citep{Schwartz2011,Hutchison2012,Orgiu2015,Ebbesen2016,Zhong2017,FeiGalGar2018}.
In particular, a strong coupling of molecular states to microcavity
photons has been demonstrated \citep{Schwartz2011,Ebbesen2016}. A
cavity induced modification of photochemical landscapes and chemical
reactivity has been reported \citep{Hutchison2012,Anoop2016}, and
the influence of the cavity vacuum fields on the charge and energy
transport in molecules has been observed experimentally \citep{Orgiu2015,Zhong2017}.
These remarkable experiments at the interface between quantum optics,
condensed matter and chemical physics triggered a theoretical activity
in developing methods that would allow to treat non-relativistic electrons
and the cavity electromagnetic modes on equal footing within a common
quantum formalism \citep{Tokatly2013PRL,Ruggenthaler2014PRA,Flick2015,GalGarFei2015,Pellegrini2015PRL,KowBenMuk2016,GalGarFei2016,GalGarFei2017,Flick2017a,Flick2017b,FeiGalGar2018,Flick2018,Vendrell2018,FliNar2018,PinRivNar2018}. 

Complex electronic structure of systems used in recent experiments
requires a QED generalization of the first-principle many-body approaches
to quantitatively describe the electronic degrees of freedom. The
most common and universal frameworks of the standard electronic structure
theory are the Green functions based many-body perturbation theory
(MBPT) \citep{OniReiRub2002,Stefanucci-book}, or the equilibrium
and time-dependent density-functional theory (DFT and TDDFT) \citep{RunGro1984,TDDFT-2012,TDDFTbyUllrich}.
In the recent years both frameworks have be generalized to include
photonic degrees of freedom. A QED extension of the non-relativistic
MBPT and the Hedin equations approach to describe many-electron systems
in microcavities have been proposed in \citep{TreMil2015,MelMar2016}.
The generalization of TDDFT, known as QED-TDDFT or QEDFT, was developed
in Refs.\citep{Tokatly2013PRL,Ruggenthaler2014PRA} and the working
power of this theory was demonstrated for several explicit examples
\citep{Pellegrini2015PRL,Flick2015,Flick2017a,Flick2017b,Flick2018}.

In practice the application of many-body methods always relies on
approximations. Apparently, in constructing approximate schemes it
is desirable to fulfill as many exactly known conditions as possible.
The conditions that follow from the fundamental conservation laws,
such as the conservation laws of the number of particles and momentum,
are of special importance because of their obvious physical significance.
In the standard self-consistent MBPT the constraints imposed by the
conservation laws have been analyzed in the seminal work by Baym \citep{Baym1962}
who proposed a general recipe for constructing so called conserving
approximations (see also a recent book \citep{Stefanucci-book}).
The importance of the exact conditions, in particular those related
to the conservation laws, for DFT and TDDFT is also well recognized
\citep{TDDFT-2012,TDDFTbyUllrich}. While in the Kohn-Sham formulation
of (TD)DFT the number of particles is conserved automatically, the
conservation of momentum requires a special care. The latter can be
restated in a form of zero exchange correlation (xc) force condition
that is directly related to a harmonic potential theorem and is crucial
for constructing non-adiabatic approximations in TDDFT \citep{Dobson1994,Vignale1995a,Vignale1995b}.
A general way to derive conserving approximations in the standard
TDDFT within the optimized effective potential (OEP) approach was
proposed in Ref.\citep{Barth2005}. The author of this work extended
the concept of $\Phi$-functional to TDDFT and showed how to construct
approximate xc potentials which are guaranteed to be conserving.

In the present paper I analyze the number of particles and the momentum
conservations laws for non-relativistic many-electron systems strongly
coupled to the cavity photon modes. The exact conditions imposed by
these conservation laws on possible approximations in the QED extension
of MBPT and QED-TDDFT are derived. I demonstrate that in spite of
the momentum exchange between the electronic and photonic subsystems
the notion of conserving approximations can be introduced for many-body
approaches to the cavity-QED, provided the electron-photon coupling
is described within the dipole approximation. The coupling to the
cavity photons induces an effective electron-electron interaction
that does not depend on the distance between the electrons and apparently
violates the Newton's third law. From the first sight one can naively
conclude that the idea of $\Phi$-derivable conserving approximations
fails here as the standard proof of conservability heavily relies
on the translation invariance of the electron-electron interaction
\citep{Baym1962,Stefanucci-book}. In the present paper I show that
this naive conclusion is not correct. It turns out that the concept
of $\Phi$-derivable approximations allows for a broader class of
electron-electron interactions that include the effective interaction
mediated by the cavity photons. I demonstrate that a properly defined
$\Phi$-functional for the cavity-QED does generate conserving approximation
both for the self energy in MBPT and for the xc potential in QED-TDDFT.
The results of this work prove that a recently proposed OEP approximation
for QED-TDDFT \citep{Pellegrini2015PRL} is conserving, and suggest
ways for its future improvements.

The structure of the paper is the following. In Sec. II I discuss
general features of the many-body problem in the cavity-QED. The basic
Hamiltonian in the length gauge is derived, the effective electron-electron
interaction mediated by the cavity photons is introduced and its physical
significance is discussed. In Sec. IIC I derive the exact force balance
equation in the cavity-QED. The main result of this section is the
zero xc force condition which has to be obeyed by any approximate
theory to ensure the correct momentum balance. In Sec. III the construction
of conserving approximations in the many-body approaches to electron-photon
systems is discussed. Here the concept of $\Phi$-functional is generalized
both for the QED extension of the self-consistent MBPT and for the
QED-TDDFT. It is proved that $\Phi$-derivable approximations fulfill
the zero xc force condition despite the lack of the Newton's third
law for the electron-electron interaction mediated by the long wavelength
cavity photons. Finally, Sec. IV summarizes the main results of this
work.

\section{Many-body problem in cavity QED and the electron force balance}

\subsection{Many-body Hamiltonian for cavity QED}

In this work I consider a typical setup of a cavity/circuit-QED which
consists of a non-relativistic many-electron system (an atom, a molecule,
an atomic cluster, a quantum dot, etc.) embedded into a micro cavity
supporting a discrete set quantum transverse electro-magnetic modes.
The electrons are confined by an external potential $V({\bf r})$
and localized within a characteristic scale $\xi$ around some point
${\bf r}_{0}$ inside the cavity. Typically the size of the electronic
subsystem $\xi$ is much smaller than the wavelength $\lambda$ of
relevant cavity modes. The small parameter $\xi/\lambda\ll1$ justifies
the description of the electron-photon coupling within the dipole
approximation. Physically this means that from the point of view of
the electromagnetic degrees of freedom the electron subsystem looks
like an effective point dipole with the following polarization density
\begin{equation}
{\bf \hat{P}}({\bf r})=e{\bf \hat{R}}\delta({\bf r}-{\bf r}_{0}),\label{P-def}
\end{equation}
where ${\bf \hat{R}}=\int{\bf r}\hat{n}({\bf r})d{\bf r}$ is the
center-of-mass coordinate of the electrons and $\hat{n}({\bf r})$
is the electron density operator. Within the dipole approximation
it is convenient to describe the combined system of electrons and
the electromagnetic field using the length gauge that in the QED context
is commonly referred to as a Power-Zienau-Wooley (PZW) gauge \citep{PowZie1959,Woolley1971}.
The corresponding many-body Hamiltonian reads
\begin{equation}
\hat{H}=\hat{H}_{{\rm el}}+\hat{H}_{{\rm e-m}}\label{H-full}
\end{equation}
Here $\hat{H}_{{\rm el}}$ is the standard Hamiltonian of a non-relativistic
many-electron system
\begin{align}
\hat{H}_{{\rm el}} & =\int d{\bf r}\left[-\hat{\psi}^{\dagger}({\bf r})\frac{\nabla^{2}}{2m}\hat{\psi}({\bf r})+V({\bf r})\hat{\psi}^{\dagger}({\bf r})\hat{\psi}({\bf r})\right]\nonumber \\
 & +\frac{1}{2}\int d{\bf r}d{\bf r}'W_{C}({\bf r}-{\bf r}')\hat{\psi}^{\dagger}({\bf r})\hat{\psi}^{\dagger}({\bf r}')\hat{\psi}({\bf r}')\hat{\psi}({\bf r}),\label{H-el}
\end{align}
where $\hat{\text{\ensuremath{\psi}}}({\bf r})$ is the fermionic
field operator, and $W_{C}({\bf r})$ is the electron-electron Coulomb
interaction potential. The second term in the Hamiltonian \eqref{H-full}
corresponds to the energy of the transverse electromagnetic field,
which also includes the dipole interaction with the electronic subsystem,
\begin{equation}
\hat{H}_{{\rm e-m}}=\frac{1}{8\pi}\int d{\bf r}\left[\hat{{\bf B}}^{2}+\left(\hat{{\bf D}}^{\perp}-4\pi\hat{{\bf P}}^{\perp}\right)^{2}\right],\label{H-em-gen}
\end{equation}
where $\hat{{\bf B}}$ is the magnetic field, and the electric field
$\hat{{\bf E}}^{\perp}=\hat{{\bf D}}^{\perp}-4\pi\hat{{\bf P}}^{\perp}$
is expressed in terms of the electric displacement $\hat{{\bf D}}^{\perp}({\bf r})$
and the transverse part $\hat{{\bf P}}^{\perp}({\bf r})$ of the electronic
polarization density of Eq. \eqref{P-def} (note that only the transverse
part of the vector field ${\bf P}({\bf r})$ is coupled to the cavity
modes). The electric displacement $\hat{{\bf D}}^{\perp}({\bf r})$
is the proper canonical variable conjugated to the magnetic filed
$\hat{{\bf B}}$. The corresponding commutation relations read as
follows
\begin{equation}
[\hat{B}_{i}({\bf r}),\hat{D}_{j}^{\perp}({\bf r}')]=-i4\pi c\varepsilon_{ijk}\partial_{k}\delta({\bf r}-{\bf r}')\label{commut-BD}
\end{equation}
where $c$ is the speed of light. One can easily check that the Heisenberg
equations of motion generated by the above commutation relations and
the Hamiltonian of Eq. \eqref{H-em-gen} indeed correctly reproduce
the Maxwell equations. 

The last step towards the basic cavity QED Hamiltonian is to introduce
a set $\{{\bf E}_{\alpha}\}$ of cavity modes labeled by the mode
index $\alpha$, and characterized by the mode's frequencies $\omega_{\alpha}$
and electric fields ${\bf E}_{\alpha}({\bf r})$. After projecting
on the cavity modes all transverse fields entering the electromagnetic
Hamiltonian \eqref{H-em-gen} one can reduce it to the following form
\begin{equation}
\hat{H}_{{\rm e-m}}=\frac{1}{2}\sum_{\alpha}\left[\hat{p}_{\alpha}^{2}+\omega_{\alpha}^{2}\left(\hat{q}_{\alpha}-\frac{\bm{\lambda}_{\alpha}}{\omega_{\alpha}}\hat{{\bf R}}\right)^{2}\right],\label{H-em-pq}
\end{equation}
where the canonical momenta $\hat{p}_{\alpha}$ and coordinates $\hat{q}_{\alpha}$
obey the standard commutation relations $[\hat{p}_{\alpha},\hat{q}_{\beta}]=-i\delta_{\alpha\beta}$,
and the vector coupling constant $\bm{\lambda}_{\alpha}=e\sqrt{4\pi}{\bf E}_{\alpha}({\bf r}_{0})$
is determined by the electric field of the $\alpha$-mode at the location
of the electronic system. Formally the Hamiltonian of Eq. \eqref{H-em-pq}
corresponds to that of a set of shifted quantum harmonic oscillators
with coordinates counted from the center-of-mass position of the electrons.
This dynamical shift is responsible for the electron-photon coupling.
By comparing the representations of Eqs. \eqref{H-em-gen} and \eqref{H-em-pq}
we easily identify the physical significance of the canonical variables
$\hat{p}_{\alpha}$ and $\hat{q}_{\alpha}$. Namely, $\sqrt{4\pi}\hat{p}_{\alpha}$
and $\sqrt{4\pi}\omega_{\alpha}\hat{q}_{\alpha}$ correspond, respectively,
to quantum amplitudes of the magnetic and the electric displacement
fields in the cavity mode $\alpha$. 

The total Hamiltonian defined by Eqs. \eqref{H-full}, \eqref{H-el},
and \eqref{H-em-pq} serves as a common starting point for the first
principle theories of realistic many-electron systems in quantum cavities
\citep{Tokatly2013PRL,Ruggenthaler2014PRA,Pellegrini2015PRL,KowBenMuk2016,Vendrell2018,PinRivNar2018,Flick2017a}.
A more detailed simple derivation of this Hamiltonian can be found
in a recent paper \citep{AbeKhoTok2018EPJB} or, in somewhat different
notations, in standard quantum optics textbooks (see e.g. Ref. \citep{MultiphotonbyFaisal}).
In the following I will use this Hamiltonian to analyze basic conservations
laws and their implications for constructing approximations.

\subsection{Electron-electron interaction induced by cavity photons}

The coupling of electrons to quantum electromagnetic modes, originating
from the second (electric energy) term in Eq. \eqref{H-em-pq}, induces
an additional electron-electron interaction via the exchange by cavity
photons. A very special harmonic form of this coupling has a deep
physical meaning that can be revealed by analyzing the structure of
the induced interaction between the electrons. Let us write more explicitly
the part of $\hat{H}_{{\rm e-m}}$ in Eq. \eqref{H-em-pq} which depends
on electronic variables
\begin{align}
\hat{H}_{{\rm int}}^{{\rm e-ph}} & =\sum_{\alpha}\Big[-\int d{\bf r}\omega_{\alpha}(\bm{\lambda}_{\alpha}{\bf r})\hat{q}_{\alpha}\hat{\psi}^{\dagger}({\bf r})\hat{\text{\ensuremath{\psi}}}({\bf r})\nonumber \\
 & +\frac{1}{2}\int d{\bf r}d{\bf r}'(\bm{\lambda}_{\alpha}{\bf r})(\bm{\lambda}_{\alpha}{\bf r}')\hat{n}({\bf r})\hat{n}({\bf r}')\Big].\label{H-int}
\end{align}
By representing the canonical mode coordinate $\hat{q}_{\alpha}$
in terms of the bosonic creation and annihilation operators $\hat{q}_{\alpha}=\frac{1}{\sqrt{2\omega_{\alpha}}}(\hat{a}_{\alpha}^{\dagger}+\hat{a}_{\alpha})$
we recognize the first term in this equation as a typical fermion-boson
coupling similar, for example, to the electron-phonon coupling in
solids. This term generates an effective retarded interaction between
the electrons. The second term in Eq. \eqref{H-int} comes from the
${\bf P}^{2}$ term in the electric energy and corresponds to an additional
instantaneous electron-electron interaction with a bilinear potential
$v({\bf r},{\bf r}')=(\bm{\lambda}_{\alpha}{\bf r})(\bm{\lambda}_{\alpha}{\bf r}')$.
The total correction to the interaction induced by the $\alpha$-mode
is a sum of the above two contributions
\begin{align}
{\cal W}_{\alpha}^{{\rm ph}}({\bf r},t;{\bf r}',t') & =\bm{\lambda}_{\alpha}{\bf r}\left[\omega_{\alpha}^{2}\langle\hat{q}_{\alpha}(t)\hat{q}_{\alpha}(t')\rangle+\delta(t-t')\right]\bm{\lambda}_{\alpha}{\bf r}'\nonumber \\
 & \equiv\bm{\lambda}_{\alpha}{\bf r}{\cal D}_{\alpha}(t-t')\bm{\lambda}_{\alpha}{\bf r}'.\label{W-ph}
\end{align}
Importantly, the two seemingly different contributions to ${\cal W}_{\alpha}^{{\rm ph}}$
enter the photon induced interaction in a special ``balanced'' way
because the coefficients in front of the corresponding terms in $\hat{H}_{{\rm int}}^{{\rm e-ph}}$
of Eq. \eqref{H-int} reflect the harmonic form of the electric energy
in Eq. \eqref{H-em-pq}. Physical implications of this balance are
most easily visible in the frequency domain. By using the standard
expression for the boson propagator $\langle\hat{q}_{\alpha}\cdot\hat{q}_{\alpha}\rangle_{\omega}=1/(\omega^{2}-\omega_{\alpha}^{2})$
one finds for the Fourier component of the function ${\cal D}_{\text{\ensuremath{\alpha}}}(t)$
in Eq. \eqref{W-ph}:
\begin{equation}
{\cal D}_{\alpha}(\omega)=\frac{\omega_{\alpha}^{2}}{\omega^{2}-\omega_{\alpha}^{2}}+1=\frac{\omega^{2}}{\omega^{2}-\omega_{\alpha}^{2}}\label{D}
\end{equation}
The first term in this equation, or, equivalently, in Eq. \eqref{W-ph},
is the displacement propagator $\langle\hat{{\bf D}}^{\perp}\cdot\hat{{\bf D}}^{\perp}\rangle_{\omega}$,
while the total induced interaction, given by the sum of the two contribution,
is nothing, but the propagator of the transverse electric field $\langle\hat{{\bf E}}^{\perp}\cdot\hat{{\bf E}}^{\perp}\rangle_{\omega}$.
This is very natural physically as the electric field is the object
that determines the energy and the force acting on charged particles.
The total propagator ${\cal D}_{\text{\ensuremath{\alpha}}}(\omega)$
in Eq. \eqref{D} is proportional to $\omega^{2}$, which reflects
a well known fact that only accelerated electron can emit radiation
felt by another electron. The corresponding electron-electron interaction
is mediated by the electric field as one would expect physically.
These fundamental physical consequences are formally related to a
balanced nature of the two interaction terms in Eq. \eqref{H-int}
as they both originate from the ${\bf E}^{2}$ contribution to the
energy of the electromagnetic field. 

An important practical outcome of the above analysis is that in any
approximate approach (unavoidable in practice) the two cavity induced
interaction terms in the Hamiltonian should be treated coherently
at the same level of approximation. Otherwise there is a danger to
violate the fundamental physics of the Maxwell electrodynamics.

Another important feature of the cavity induced interaction Eq. \eqref{W-ph}
is its dependence on the spatial coordinates of the interacting particles.
In contrast to the direct Coulomb interaction the function ${\cal W}_{\alpha}^{{\rm ph}}({\bf r};{\bf r}')$
is not translation invariant, that is, it does not depend on the coordinate
difference ${\bf r}-{\bf r}'$. This implies the lack of the Newton's
third law, and therefore one may expect a net force exerted on the
center-of-mass of the electronic system due to the photon induced
interaction. This is of course no surprising as the coupling to cavity
photons can produce a net force on the electrons leading, for example,
to a radiation friction. 

The absence of translation invariance of the electron-electron interaction
also has serious technical consequences for the demonstration of conservability
of $\Phi$-derivable approximations. As the classical argumentation
by Baym heavily relies on the fact that the interaction potential
depends on the distance between particles \citep{Baym1962,Stefanucci-book}
the usual conservability proof apparently fails in the presence of
cavity photons. These points will be carefully analyzed in subsequent
sections.

\subsection{Dynamics of observables and the force balance\label{sub:Dynamics-of-observables}}

In the present context the following physical observables are of interest:
(i) the electron density $n({\bf r},t)=\langle\hat{n}({\bf r})\rangle$,
(ii) the electron current ${\bf j}({\bf r},t)=\langle\hat{{\bf j}}({\bf r})\rangle$,
and (iii) the expectation value of the electric displacement amplitude
$Q_{\alpha}(t)=\langle\hat{q}_{\alpha}\rangle$. The dynamics of the
observables is governed by the corresponding Heisenberg equations
of motion,
\begin{align}
\partial_{t}\langle\hat{q}_{\alpha}\rangle & =i\langle[\hat{H},\hat{q}_{\alpha}]\rangle;\quad\partial_{t}\langle\hat{p}_{\alpha}\rangle=i\langle[\hat{H},\hat{p}_{\alpha}]\rangle,\label{p-q-dynamics}\\
 & \partial_{t}\langle\hat{n}({\bf r})\rangle=i\langle[\hat{H},\hat{n}({\bf r})]\rangle,\label{n-dynamics}\\
 & \partial_{t}\langle\hat{{\bf j}}({\bf r})\rangle=i\langle[\hat{H},\hat{{\bf j}}({\bf r})]\rangle.\label{j-dynamics}
\end{align}
The couple of equations in Eq. \eqref{p-q-dynamics} corresponds to
the mode-projected Maxwell equations for the expectation values of
the transverse fields. After evaluating the commutators and eliminating
the magnetic amplitude $\langle\hat{p}_{\alpha}\rangle$, we obtain
the following projected ``wave equation'' for the electric displacement
\begin{equation}
\partial_{t}^{2}Q_{\alpha}+\omega_{\alpha}^{2}Q_{\alpha}-\omega_{\alpha}\bm{\lambda}_{\alpha}{\bf R}=0,\label{Maxwell}
\end{equation}
where ${\bf R}(t)=\langle\hat{\text{{\bf R}}}\rangle=\int{\bf r}n({\bf r},t)d{\bf r}$
is the expectation value of the center-of-mass coordinate of the electrons. 

Since the electron density operator commutes with $\hat{H}_{{\rm e-m}}$
of Eq. \eqref{H-em-pq} the equation of motion Eq. \eqref{n-dynamics}
reduces to the standard continuity equation which reflects the local
conservation low of the number of electrons and stays unmodified by
the presence of the cavity
\begin{equation}
\partial_{t}n+\nabla{\bf j}=0.\label{continuity}
\end{equation}
In contrast, for the electron current the commutator $[\hat{H}_{{\rm e-m}},\hat{{\bf j}}({\bf r})]\ne0$
does not vanish thus producing a force exerted on electrons from the
photonic subsystem. This force is our main concern here. Equation
\eqref{j-dynamics} describes a local electron force balance and can
be written more explicitly as follows,
\begin{equation}
\partial_{t}{\bf j}-{\bf F}^{{\rm str}}-\sum_{\alpha}{\bf f}^{\alpha}+n\nabla V=0.\label{local-force}
\end{equation}
Here the last term is the force density due to the external classical
potential (the second term in $\hat{H}_{{\rm el}}$ of Eq. \eqref{H-el}).
The second term in Eq. \eqref{local-force} is the electron stress
force originating from the kinetic $\hat{T}$ and the interaction
$\hat{W}_{C}$ contributions in $\hat{H}_{{\rm el}}$,
\begin{equation}
F_{k}^{{\rm str}}({\bf r},t)=i\langle[\hat{T}+\hat{W}_{C},\hat{j}_{k}({\bf r})]\rangle=-\partial_{i}\Pi_{ik}\label{F-str}
\end{equation}
Since the Hamiltonian $\hat{T}+\hat{W}_{C}$ is translation invariant
the local stress force obeys the Newton's third law. This means that
the vector ${\bf F}^{{\rm str}}({\bf r},t)$ can be represented as
a divergence of a second rank tensor $\Pi_{ik}({\bf r},t)$ -- the
electron stress tensor \citep{Tokatly2005PRBa}. Finally, the third
term is the force due to the coupling to the cavity modes,
\begin{equation}
{\bf f}^{\alpha}({\bf r},t)=\langle\bm{\lambda}_{\alpha}(\omega_{\alpha}\hat{q}_{\alpha}-\bm{\lambda}_{\alpha}\cdot\hat{{\bf R}})\hat{n}({\bf r})\rangle=\langle\hat{{\bf e}}_{\alpha}\;\hat{n}({\bf r})\rangle\label{f-alpha}
\end{equation}
where we recognize $\hat{{\bf e}}_{\alpha}=\bm{\lambda}_{\alpha}(\omega_{\alpha}\hat{q}_{\alpha}-\bm{\lambda}_{\alpha}\cdot\hat{{\bf R}})$
as an operator of the $\alpha$-mode electric field at the position
of the electronic system. Not surprisingly the force density Eq. \eqref{f-alpha}
produced by the photonic subsystem is given by the equal time correlation
function of the electric field and the electron density operators.

Equations of the global electron force/momentum balance is obtained
by integrating Eq. \eqref{local-force} over the space variable ${\bf r}$.
Because of the Newton's third law the net electronic stress force
vanishes, $\int{\bf F}^{{\rm str}}({\bf r},t)d{\bf r}=0,$ and we
are left with the following result
\begin{equation}
\partial_{t}{\bf P}=\sum_{\alpha}\bm{\lambda}_{\alpha}(\omega_{\alpha}Q_{\alpha}-\bm{\lambda}_{\alpha}\cdot{\bf R})N-\int n\nabla Vd{\bf r},\label{global-force}
\end{equation}
where ${\bf P}(t)=\int{\bf j}({\bf r},t)d{\bf r}$ is the total momentum
of the electrons. The first term in the right hand side of Eq. \eqref{global-force}
is the net force exerted on electrons from the cavity photons,
\begin{equation}
\int{\bf f}^{\alpha}({\bf r},t)d{\bf r}=\bm{\lambda}_{\alpha}\big[\omega_{\alpha}Q_{\alpha}-\bm{\lambda}_{\alpha}\cdot{\bf R}\big]N=\langle\hat{{\bf e}}_{\alpha}\rangle N,\label{f-alfpha-net}
\end{equation}
which is determined by the expectation (mean) value of the electric
field operator. Having in mind this result I represent the local force
${\bf f}^{\alpha}({\bf r},t)$ as sum of a mean field and an exchange-correlation
(xc) contributions
\[
{\bf f}^{\alpha}({\bf r},t)={\bf f}_{{\rm mf}}^{\alpha}({\bf r},t)+{\bf f}_{{\rm xc}}^{\alpha}({\bf r},t),
\]
where the mean field force is given by the product of the expectation
values of the electric field and the density,
\begin{equation}
{\bf f}_{{\rm mf}}^{\alpha}({\bf r},t)=\langle\hat{{\bf e}}_{\alpha}\rangle n({\bf r},t)=\bm{\lambda}_{\alpha}\big[\omega_{\alpha}Q_{\alpha}(t)-\bm{\lambda}_{\alpha}{\bf R}(t)\big]n({\bf r},t),\label{f-alpha-mf}
\end{equation}
while the xc force is determined by the equal time correlation function
of the fluctuation operators,
\begin{equation}
{\bf f}_{{\rm xc}}^{\alpha}({\bf r},t)=\langle\Delta\hat{{\bf e}}_{\alpha}\Delta\hat{n}({\bf r})\rangle=\langle\bm{\lambda}_{\alpha}(\omega_{\alpha}\Delta\hat{q}_{\alpha}-\bm{\lambda}_{\alpha}\Delta\hat{{\bf R}})\Delta\hat{n}({\bf r})\rangle\label{f-alpha-xc}
\end{equation}
Here the fluctuation operators are defined in a standard manner as
$\Delta\hat{O}=\hat{O}-\langle\hat{O}\rangle$.

Now the most important outcome of the above analysis can be formulated
as follows. Within the dipole approximations the exact global force
exerted on the electrons from the cavity photons is exhausted by the
mean field contribution, 
\[
\int{\bf f}^{\alpha}({\bf r},t)d{\bf r}=\int{\bf f}_{{\rm mf}}^{\alpha}({\bf r},t)d{\bf r}.
\]
In other words, the correct force balance of Eq. \eqref{global-force}
is guaranteed only if the global xc force from the photons vanishes,
\begin{equation}
\int{\bf f}_{{\rm xc}}^{\alpha}({\bf r},t)d{\bf r}=\int\langle\Delta\hat{{\bf e}}_{\alpha}\Delta\hat{n}({\bf r})\rangle d{\bf r}=0.\label{zero-xc-force}
\end{equation}
This condition generalized the requirement of the momentum conservation
to systems of electrons coupled to long wavelength cavity photons.
Obviously, it is desirable for approximate many-body theories to fulfill
the above exact condition. The corresponding approximations can be
naturally called conserving.

In the next section I consider two possible first principle approaches
to the non-equilibrium many-body theory: (i) a self-consistent MBPT,
and (ii) the QED-TDDFT of Refs. \citep{Tokatly2013PRL,Ruggenthaler2014PRA}.
I will show how the standard arguments leading to conserving approximations
in the usual MBPT \citep{Baym1962,Stefanucci-book} and TDDFT \citep{Barth2005}
can be generalized to the case of electron-photon systems in the cavity-QED.

\section{Conserving approximations: generalization of baym argument}

\subsection{Self-consistent many-body perturbation theory\label{sub:Self-consistent-many-body-theory}}

Let us start from a field theoretical formulation of the many-body
problem -- the MBPT. The key object of this approach is a one-particle
Green function $G(1,2)=-i\langle T_{{\cal C}}\psi({\bf r}_{1},t_{1})\psi^{\dagger}({\bf r}_{2},t_{2})\rangle$
where the operator $T_{{\cal C}}$ orders ``time'' arguments along
a certain contour ${\cal C}$ in a complex plane. Depending on the
choice of the contour we recover different versions of MBPT, such
as, zero-temperature, equilibrium Matsubara, or non-equilibrium Keldysh
formalisms \citep{LeeSte2013,Stefanucci-book}. As in the present
paper I am interested in dynamics the Keldysh time contour is assumed.

By explicitly separating the mean field contribution I represent the
equations of motion for the Green functions in the following form\begin{widetext}

\begin{align}
i\partial_{t_{1}}G(1,2)-\hat{h}_{{\rm mf}}(1)G(1,2)-\int d3\Sigma_{{\rm xc}}(1,3)G(3,2) & =\delta(1-2)\label{eq:G-1}\\
-i\partial_{t_{2}}G(1,2)-\hat{h}_{{\rm mf}}(2)G(1,2)-\int d3G(1,3)\Sigma_{{\rm xc}}(3,2) & =\delta(1-2)\label{eq:G-2}
\end{align}
\end{widetext} where the mean field Hamiltonian reads
\begin{equation}
\hat{h}_{{\rm mf}}({\bf r},t)=-\frac{\nabla^{2}}{2m}+V+V_{H}+\sum_{\alpha}(\omega_{\alpha}Q_{\alpha}-\bm{\lambda}_{\alpha}{\bf R})\bm{\lambda}_{\alpha}{\bf r}.\label{h-mf}
\end{equation}
Here $V_{H}({\bf r},t)=\int W_{C}({\bf r}-{\bf r}')n({\bf r}',t)d{\bf r}'$
is the usual Hartree potential, and the displacement amplitude $Q_{\alpha}(t)$
satisfies the projected Maxwell equation \eqref{Maxwell}. The xc
self energy $\Sigma_{{\rm xc}}(1,2)$ is constructed according to
the standard diagrammatic rules from the one-particle Green functions
and the total effective interaction ${\cal W}(1,2)$ that consists
of the direct Coulomb interaction and the cavity induced correction
${\cal W}^{{\rm ph}}(1,2)$ of Eq. \eqref{W-ph}: 
\begin{equation}
{\cal W}(1,2)=W_{C}({\bf r}_{1}-{\bf r}_{2})\delta(t_{1}-t_{2})+\sum_{\alpha}\bm{\lambda}_{\alpha}{\bf r}_{1}{\cal D}_{\alpha}(t_{1}-t_{2})\bm{\lambda}_{\alpha}{\bf r}_{2}.\label{W-total}
\end{equation}
Specifically, $\Sigma_{{\rm xc}}$ is given by one-particle irreducible
skeleton diagrams, excluding the Hartree diagram. The latter is represented
by the last two terms in the mean field Hamiltonian \eqref{h-mf}. 

The equations of motion for the electron density $n({\bf r},t)$ and
the total electron momentum ${\bf P}(t)$ can be now straightforwardly
derived from Eqs. \eqref{eq:G-1}-\eqref{eq:G-2} with any given $\Sigma_{{\rm xc}}$
\citep{Stefanucci-book}. By comparing the obtained equations with
their exact counterparts of Eqs. \eqref{continuity} and \eqref{global-force}
one finds the conditions which should be fulfilled by the self energy
to guarantee the correct form of the conservation laws. In particular,
the continuity equation is recovered if $\Sigma_{{\rm xc}}$ satisfies
the condition
\begin{equation}
\int d2[\Sigma_{{\rm xc}}(1,2)G(2,1)-G(1,2)\Sigma_{{\rm xc}}(2,1)]=0.\label{N-condition}
\end{equation}
Similarly we find that the correct momentum balance of Eq. \eqref{global-force}
is reproduced provided the following equation is fulfilled
\begin{equation}
\int d{\bf r}_{1}d2[\Sigma_{{\rm xc}}(1,2)\nabla_{1}G(2,1)-G(1,2)\nabla_{1}\Sigma_{{\rm xc}}(2,1)]=0.\label{P-condition}
\end{equation}
\eqref{N-condition} and \eqref{P-condition} coincide with the well
known conditions imposed on the self energy of conserving of approximations
in the standard MBPT \citep{Baym1962,Stefanucci-book}. The reason
is that at the level of the conservation laws of interest the cavity
induced modifications are exactly captured by the mean field part
of Eqs. \eqref{eq:G-1}-\eqref{eq:G-2}. Therefore, similarly to the
standard MBPT, the many-body xc corrections due to $\Sigma_{{\rm xc}}$
should not contribute to the conservation laws. Note that \eqref{P-condition}
is nothing but the statement of vanishing xc force expressed in terms
of the self energy of MBPT.

Let us now examine the standard arguments for constructing conserving
approximations for xc self energy. The common prescription relies
on the concept of $\Phi$-functional due to Baym \citep{Baym1962}.
A functional $\Phi[G]$ of the Green function $G$ is constructed
by selecting a subset of connected ``energy diagrams''. An example
of $\Phi$-functional is shown 
\begin{figure}
\includegraphics[scale=0.7]{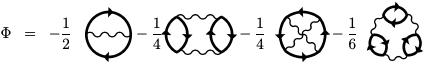}\caption{Example of $\Phi$-functional \citep{Stefanucci-book} }
\end{figure}
 on Fig. 1, where thick arrows denote the Green functions $G(1,2)$
and wiggled lines stand for the particle-particle interaction ${\cal W}(1,2)$
\citep{Stefanucci-book}. For a given $\Phi[G]$, the corresponding
self energy is defined as the following functional derivative
\begin{equation}
\Sigma_{{\rm xc}}(1,2)=\frac{\delta\Phi}{\delta G(2,1)}.\label{Sigma-def}
\end{equation}
Approximations generated via this procedure are called $\Phi$-derivable. 

To prove that a $\Phi$-derivable approximation is conserving one
has to look on symmetries of the underlying $\Phi$-functional. In
particular the condition Eq. \eqref{N-condition} is a consequence
of the gauge invariance. By construction, any diagram for $\Phi[G]$
is invariant with respect to the following replacement $G(1,2)\mapsto e^{i\Lambda(1)}G(1,2)e^{-i\Lambda(2)}$
where $\Lambda({\bf r},t)$ is an arbitrary function. By requiring
that $\Phi[G]$ is unchanged under the corresponding infinitesimal
variation, $G(1,2)\mapsto G(1,2)+\delta G(1,2)$ with $\delta G(1,2)=i[\Lambda(1)-\Lambda(2)]G(1,2)$,
and using the definition of Eq. \eqref{Sigma-def} we immediately
obtain Eq. \eqref{N-condition}. 

Obviously, the gauge invariance of the diagrams for $\Phi[G]$ does
not depend on a specific form of the particle-particle interaction
${\cal W}(1,2)$ (only the space-time locality of vertices where two
Green functions meet is important). Therefore the modification of
the interaction by the cavity photons does not influence the standard
arguments and we can safely conclude that $\Phi$-derivable approximations
still conserve the number of particles in the presence of quantum
electromagnetic field in cavity-QED. In contrast, the situation with
the momentum/force balance is very different. The standard proof of
the momentum conservation assumes that $\Phi[G]$ is unchanged under
a time-dependent shift of space arguments of the Green function $G({\bf r}_{1},t_{1};{\bf r}_{2},t_{2})\mapsto G({\bf r}_{1}+{\bf u}(t_{1}),t_{1};{\bf r}_{2}+{\bf u}(t_{2}),t_{2})$.
However, this is only true for an instantaneous translation invariant
particle-particle interaction ${\cal W}(1,2)=W({\bf r}_{1}-{\bf r}_{2})\delta(t_{1}-t_{2})$
that implies a Newton third law, $\nabla_{1}W({\bf r}_{1},{\bf r}_{2})=-\nabla_{2}W({\bf r}_{1},{\bf r}_{2})$.
The coupling to cavity modes breaks this property by producing an
additional photon-exchange interaction given by the second term in
Eq. \eqref{W-total}. Below I will show how to generalize the proof
and demonstrate that the $\Phi$-derivable approximations are consistent
with the zero xc force for a more general class of particle-particle
interactions, including the one of Eq. \eqref{W-total}.

First I notice that by construction $\Phi[G,{\cal W}]$ is a functional
of the Green function $G$ and the interaction ${\cal W}$. Another
simple observation is that, irrespectively of a specific form of interaction,
the $\Phi$-functional is unchanged if we perform a simultaneous shift
of space arguments both in $G(1,2)$ and in ${\cal W}(1,2)$,
\begin{align*}
G({\bf r}_{1},t_{1};{\bf r}_{2},t_{2})\mapsto G({\bf r}_{1}+{\bf u}(t_{1}),t_{1};{\bf r}_{2}+{\bf u}(t_{2}),t_{2}) & ,\\
{\cal W}({\bf r}_{1},t_{1};{\bf r}_{2},t_{2})\mapsto{\cal W}({\bf r}_{1}+{\bf u}(t_{1}),t_{1};{\bf r}_{2}+{\bf u}(t_{2}),t_{2}) & .
\end{align*}
The reason for the invariance is the space integration in all vertices
in each diagram for a given approximate $\Phi$-functional. Because
of this invariance a variation $\delta\Phi$ generated by an infinitesimal
translation of spatial arguments of $G$ and ${\cal W}$ should vanish,
\begin{equation}
\int d2\left[\frac{\delta\Phi[G,{\cal W}]}{\delta G(1,2)}\delta G(1,2)+\frac{\delta\Phi[G,{\cal W}]}{\delta{\cal W}(1,2)}\delta{\cal W}(1,2)\right]=0,\label{variaton-gen}
\end{equation}
where the variations of the two point functions are defined as $\delta F(1,2)={\bf u}(t_{1})\nabla_{1}F(1,2)+{\bf u}(t_{2})\nabla_{2}F(1,2)$.
The functional derivative in first term in Eq. \eqref{variaton-gen}
is by definition the xc self energy $\Sigma_{{\rm xc}}$ of Eq. \eqref{Sigma-def}.
By a direct inspection of diagrams the functional derivative in the
second terms is easily recognized as the density response function
$\chi(1,2)$ \citep{AlmBarLee1999,Stefanucci-book},
\begin{equation}
\frac{\delta\Phi[G,{\cal W}]}{\delta{\cal W}(1,2)}=-\frac{1}{2}\chi(1,2).\label{chi-def}
\end{equation}
Using the above identification of the functional derivatives one can
rewrite the identity of Eq. \eqref{variaton-gen} in the following
form
\begin{align}
\int d{\bf r}_{1}d2\left[\Sigma_{{\rm xc}}(1,2)\nabla_{1}G(2,1)-G(1,2)\nabla_{1}\Sigma_{{\rm xc}}(2,1)\right]\nonumber \\
=\frac{1}{2}\int\int d{\bf r}_{1}d2\left[\chi(1,2)\nabla_{1}{\cal W}(2,1)+\nabla_{1}{\cal W}(1,2)\chi(2,1)\right]\label{identity}
\end{align}
where the left and the right hand sides correspond, respectively,
to the first and the second terms in Eq. \eqref{variaton-gen}. The
left hand side in Eq. \eqref{identity} coincides with the left hand
side in Eq. \eqref{P-condition}. Hence the zero xc force condition
is satisfied if the right hand side in the above Eq. \eqref{identity}
vanishes. It obviously vanishes for an instantaneous translation invariant
interaction of the form ${\cal W}(1,2)=W({\bf r}_{1}-{\bf r}_{2})\delta(t_{1}-t_{2})$
which satisfies the Newton's third law. There is however another possibility
for the interaction to nullify the right hand side in Eq. \eqref{identity}.
Notice that because of the gauge invariance the response function
$\chi(1,2)$ satisfies the following identity $\int d{\bf r}_{1}\chi(1,2)=\int d{\bf r}_{2}\chi(1,2)=0$,
which guarantees the absence of the density response generated by
a spatially uniform scalar potential. Using this property we find
that the right hand side in Eq. \eqref{identity} also vanishes for
a biliner interaction of the form ${\cal W}(1,2)=r_{1}^{i}{\cal D}^{ij}(t_{1},t_{2})r_{2}^{j}$,
where ${\cal D}^{ij}(t_{1},t_{2})$ is an arbitrary function of only
time variables. Therefore a generic particle-particle interaction
consistent with the zero xc force condition of Eq. \eqref{P-condition}
is the following,
\begin{equation}
{\cal W}(1,2)=W({\bf r}_{1}-{\bf r}_{2})\delta(t_{1}-t_{2})+r_{1}^{i}{\cal D}^{ij}(t_{1},t_{2})r_{2}^{j}.\label{W-conserving}
\end{equation}
This is exactly the form of the particle-particle interaction we have
found for the cavity-QED, see Eq. \eqref{W-total}. Specifically for
the many-electron system interacting with long wavelength cavity photons
$W({\bf r}_{1}-{\bf r}_{2})=W_{C}(|{\bf r}_{1}-{\bf r}_{2}|)$ is
the Coulomb interaction potential, and ${\cal D}^{ij}(t_{1},t_{2})=\sum_{\alpha}\lambda_{\alpha}^{i}{\cal D}_{\alpha}(t_{1}-t_{2})\lambda_{\alpha}^{j}$
is the electric field propagator for the cavity photons.

The most important conclusion is that despite the coupling to quantum
cavity modes breaks the Newton's third law, all $\Phi$-derivable
approximations are still both number- and momentum-conserving.

\subsection{Time-dependent density functional theory}

In this subsection the previously obtained results will be applied
to the construction of conserving approximations for xc potential
in the QED extension of TDDFT. 

I start with a brief review of QED-TDDFT in a form proposed in Ref.
\citep{Tokatly2013PRL} and further elaborated in Ref. \citep{Ruggenthaler2014PRA}.
Generically QED-TDDFT relies on the following mapping theorem \citep{Tokatly2013PRL}.
The time-dependent many-body wave function $|\Psi(t)\rangle$ of the
electron-photon system and the external one-particle potential $V({\bf r},t)$
are unique functionals of the initial state $|\Psi_{0}\rangle$, the
electron density $n({\bf r},t)$ and the expectation values $Q_{\alpha}(t)$
of the displacement amplitudes. This statement allows us to calculate
the basics observables, $n({\bf r},t)$ and $Q_{\alpha}(t)$, by solving
a system of self-consistent Kohn-Sham-Maxwell equations for a set
of one-particle KS orbitals $\phi_{i}({\bf r},t)$ and the displacement
amplitudes $Q_{\alpha}(t)$: 
\begin{align}
i\partial_{t}\phi_{i} & =-\frac{\nabla^{2}}{2m}\phi_{i}+\Big[V_{S}+\sum_{\alpha}(\omega_{\alpha}Q_{\alpha}-\bm{\lambda_{\alpha}}{\bf R})\bm{\lambda}_{\alpha}{\bf r}\Big]\phi_{i}\label{KS-equations}\\
 & \partial_{t}^{2}Q_{\alpha}+\omega_{\alpha}^{2}Q_{\alpha}-\omega_{\alpha}\bm{\lambda}_{\alpha}{\bf R}=0.\label{Maxwell-2}
\end{align}
Here the KS potential $V_{S}({\bf r},t)$ is a sum of the external
potential $V({\bf r},t)$, the Hartree potential $V_{H}({\bf r},t)$,
and the xc potential $V_{{\rm xc}}({\bf r},t)$, 
\begin{equation}
V_{S}({\bf r},t)=V({\bf r},t)+V_{H}[n]({\bf r},t)+V_{{\rm xc}}[n,Q]({\bf r},t).\label{V-S}
\end{equation}
The xc potential is a functional of the basic observables $V_{{\rm xc}}[n,Q]$
which encodes all complicated many-body effects. It is adjusted in
such a way that exact electron density is reproduced in the system
of noninteracting KS particles, $n({\bf r},t)=\sum_{\alpha=1}^{N}|\phi_{i}({\bf r},t)|^{2}$.
In has been shown in Ref. \citep{Tokatly2013PRL} that similarly to
the usual TDDFT \citep{TDDFT-2012,TDDFTbyUllrich} the xc potential
of QED-TDDFT satisfies the zero force condition
\begin{equation}
\int n({\bf r},t)\nabla V_{{\rm xc}}({\bf r},t)d{\bf r}=0.\label{zero-force}
\end{equation}
Clearly this condition is a direct consequence of the conservations
laws derived in Sec.\ref{sub:Dynamics-of-observables}. In fact, Eq.
\eqref{zero-force} ensures the correct global momentum balance in
the coupled electron-photon system. It is worth noting that in the
KS formulation of any TDDFT the continuity equation is satisfied automatically.

Any practical application of TDDFT requires approximations for the
xc potential. In the standard TDDFT a general scheme of constructing
conserving optimized effective potential (OEP) approximations for
$V_{{\rm xc}}$ has been proposed in Ref. \citep{Barth2005}. I will
show that this scheme can be easily adopted to QED-TDDFT and prove
that here it also produces conserving xc potentials.

Following the idea of Ref. \citep{Barth2005} I consider an approximate
$\Phi$-functional of MBPT, but evaluate it at the KS Green function
$\Phi[G_{s},{\cal W}]$, where the KS Green function is the one-particle
propagator related to the KS Hamiltonian in Eq. \eqref{KS-equations}.
Because of the density-potential mapping $G_{s}[n]$ is a functional
of the electron density. Therefore the above $\Phi$-functional can
be also regarded as a functional of the density and the particle-particle
interaction, $\Phi[G_{s},{\cal W}]=\tilde{\Phi}[n,{\cal W}]$. Importantly,
the $\Phi$-functional depends on $n$ only via $G_{s}$. The xc potential
is now defined as follows \citep{Barth2005}
\begin{equation}
V_{{\rm xc}}({\bf r},t)=\frac{\delta\tilde{\Phi}[n,{\cal W}]}{\delta n({\bf r},t)}.\label{Vxc-def}
\end{equation}
The level of OEP approximation in this scheme depends on the diagrams
taken into account in $\Phi[G_{s},{\cal W}]$.

The first step in proving that xc potential of Eq. \eqref{Vxc-def}
is conserving is to analyze the symmetry of the functional $\tilde{\Phi}[n,{\cal W}]$.
Let us shift the spatial argument of the density by a time-dependent
amount $n({\bf r}+{\bf u}(t),t)$. This will generate the corresponding
shift of arguments in the KS Green function $G_{s}({\bf r}_{1},t_{1};{\bf r}_{2},t_{2})\mapsto G_{s}({\bf r}_{1}+{\bf u}(t_{1}),t_{1};{\bf r}_{2}+{\bf u}(t_{2}),t_{2})$.
If we simultaneously perform a similar shift in the particle-particle
interaction ${\cal W}({\bf r}_{1},t_{1};{\bf r}_{2},t_{2})\mapsto{\cal W}({\bf r}_{1}+{\bf u}(t_{1}),t_{1};{\bf r}_{2}+{\bf u}(t_{2}),t_{2})$,
then, in full analogy with the discussion in Sec.\ref{sub:Self-consistent-many-body-theory},
the $\Phi$-functional will remain unchanged. By requiring the invariance
of $\Phi[G_{s},{\cal W}]=\tilde{\Phi}[n,{\cal W}]$ with respect to
the infinitesimal version of the above shift and performing calculations
similar to those in the previous section we arrive at the following
identity,
\begin{align}
 & \qquad-\int n({\bf r}_{1},t_{1})\nabla V_{{\rm xc}}({\bf r}_{1},t_{1})d{\bf r}_{1}\label{TDDFT-identity}\\
= & \frac{1}{2}\int\int d{\bf r}_{1}d2\left[\tilde{\chi}(1,2)\nabla_{1}{\cal W}(2,1)+\nabla_{1}{\cal W}(1,2)\tilde{\chi}(2,1)\right].\nonumber 
\end{align}
Here $\tilde{\chi}(2,1)$ is defined similarly to Eq.\eqref{chi-def},
but with the $\Phi$-functional evaluated at the KS Green function,
\begin{equation}
\frac{\delta\Phi[G_{s},{\cal W}]}{\delta{\cal W}(1,2)}=-\frac{1}{2}\tilde{\chi}(1,2).\label{chi-tilde}
\end{equation}
The function $\tilde{\chi}(2,1)$ is not the density response function
of our physical system. However, by construction it is given by the
density response diagrams constructed from the physical interaction
${\cal W}$ and the KS Green $G_{s}$ that is a legitimate one-particle
propagator. Therefore $\tilde{\chi}(2,1)$ obeys all fundamental properties
of the density response function, in particular $\int d{\bf r}_{1}\tilde{\chi}(1,2)=\int d{\bf r}_{2}\tilde{\chi}(1,2)=0$.
Therefore using the same reasoning as in Sec.\ref{sub:Self-consistent-many-body-theory}
we conclude that the right hand side in Eq.\eqref{TDDFT-identity}
vanishes if the particle-particle interaction has a generic form of
Eq.\eqref{W-conserving}. In other words I have demonstrated that
the described cavity-QED generalization of the OEP construction generates
conserving approximations for the xc potential.

Recently an OEP approximation based on the first order xc self energy
has been proposed for QED-TDDFT \citep{Pellegrini2015PRL}. A good
performance of this approximation has been already demonstrated in
several publications \citep{Pellegrini2015PRL,Flick2017a,Flick2018},
however it remained unclear whether it satisfies the fundamental zero
force theorem. In terms of the $\Phi$-functional the OEP $V_{{\rm xc}}$
of Ref.\citep{Pellegrini2015PRL} is generated by the first diagram
on Fig.1. Hence the results of the present section imply that this
approximation is perfectly conserving.

\section{conclusion}

In conclusion, by considering the many-body problem for electronic
systems strongly coupled to the cavity photon modes I derived the
electron force balance equation, and analyzed the exact conditions
imposed by this equation on approximate many-body approaches to the
cavity-QED. The correct momentum balance in the combined system is
guarantied if a properly defined global xc force exerted on electrons
from the photonic subsystem vanishes. This condition is similar to
the momentum conservability in the standard many-body theory. To construct
approximations which fulfill the zero xc force constraint in the frameworks
of MBPT and OEP QED-TDDFT I generalized the concepts of $\Phi$-functional
and $\Phi$-derivable approximations. In the case of cavity-QED the
conservability of $\Phi$-derivable approximations is not as trivial
as it may appear on the first sight. The reason is that the exchange
by the long wavelength cavity photons induces an effective electron-electron
interaction violating the Newton's third law, which can be traced
back to the momentum transfer between the electronic and photonic
subsystems. Nonetheless, the concept of conserving approximations
can be introduced and all $\Phi$-derivable approximations remain
conserving as long as the dipole approximation is valid for the electron-photon
coupling (which is the case in most experimentally relevant situations).
In particular, this result implies that the recently proposed first
order OEP xc potential for QED-TDDFT \citep{Pellegrini2015PRL} is
conserving. 

An interesting observation is that $\Phi$-derivable approximations
are conserving independently on the specific form of the electric
field propagator ${\cal D}^{ij}(t_{1},t_{2})$ in Eq.\eqref{W-conserving}.
This suggests a natural and simple way to improve/generalize the OEP
of Ref.\citep{Pellegrini2015PRL} without introducing extra numerical
complexity. In the genuine first order OEP one uses the bare photon
propagator in the effective interaction, which obviously misses the
renormalization of the cavity photons. However the photon renormalization
effects can be easily mimicked without breaking the momentum balance
by replacing the bare propagator with an effective one constructed
phenomenologically on physical grounds, or imported form a simplified
solvable system. It would be interesting to explore this possibility
in the future. 
\begin{acknowledgments}
This work is supported by the Spanish Ministerio de Economía y Competividad
(MINECO) Project No. FIS2016-79464-P and by the ``Grupos Consolidados
UPV/EHU del Gobierno Vasco'' (Grant No. IT578-13). 
\end{acknowledgments}

\bibliographystyle{apsrev4-1}

\begin{thebibliography}{49}%
\makeatletter
\providecommand \@ifxundefined [1]{%
 \@ifx{#1\undefined}
}%
\providecommand \@ifnum [1]{%
 \ifnum #1\expandafter \@firstoftwo
 \else \expandafter \@secondoftwo
 \fi
}%
\providecommand \@ifx [1]{%
 \ifx #1\expandafter \@firstoftwo
 \else \expandafter \@secondoftwo
 \fi
}%
\providecommand \natexlab [1]{#1}%
\providecommand \enquote  [1]{``#1''}%
\providecommand \bibnamefont  [1]{#1}%
\providecommand \bibfnamefont [1]{#1}%
\providecommand \citenamefont [1]{#1}%
\providecommand \href@noop [0]{\@secondoftwo}%
\providecommand \href [0]{\begingroup \@sanitize@url \@href}%
\providecommand \@href[1]{\@@startlink{#1}\@@href}%
\providecommand \@@href[1]{\endgroup#1\@@endlink}%
\providecommand \@sanitize@url [0]{\catcode `\\12\catcode `\$12\catcode
  `\&12\catcode `\#12\catcode `\^12\catcode `\_12\catcode `\%12\relax}%
\providecommand \@@startlink[1]{}%
\providecommand \@@endlink[0]{}%
\providecommand \url  [0]{\begingroup\@sanitize@url \@url }%
\providecommand \@url [1]{\endgroup\@href {#1}{\urlprefix }}%
\providecommand \urlprefix  [0]{URL }%
\providecommand \Eprint [0]{\href }%
\providecommand \doibase [0]{http://dx.doi.org/}%
\providecommand \selectlanguage [0]{\@gobble}%
\providecommand \bibinfo  [0]{\@secondoftwo}%
\providecommand \bibfield  [0]{\@secondoftwo}%
\providecommand \translation [1]{[#1]}%
\providecommand \BibitemOpen [0]{}%
\providecommand \bibitemStop [0]{}%
\providecommand \bibitemNoStop [0]{.\EOS\space}%
\providecommand \EOS [0]{\spacefactor3000\relax}%
\providecommand \BibitemShut  [1]{\csname bibitem#1\endcsname}%
\let\auto@bib@innerbib\@empty
\bibitem [{\citenamefont {Raimond}\ \emph {et~al.}(2001)\citenamefont
  {Raimond}, \citenamefont {Brune},\ and\ \citenamefont
  {Haroche}}]{RaiBruHar2001}%
  \BibitemOpen
  \bibfield  {author} {\bibinfo {author} {\bibfnamefont {J.~M.}\ \bibnamefont
  {Raimond}}, \bibinfo {author} {\bibfnamefont {M.}~\bibnamefont {Brune}}, \
  and\ \bibinfo {author} {\bibfnamefont {S.}~\bibnamefont {Haroche}},\ }\href
  {\doibase 10.1103/RevModPhys.73.565} {\bibfield  {journal} {\bibinfo
  {journal} {Rev. Mod. Phys.}\ }\textbf {\bibinfo {volume} {73}},\ \bibinfo
  {pages} {565} (\bibinfo {year} {2001})}\BibitemShut {NoStop}%
\bibitem [{\citenamefont {Mabuchi}\ and\ \citenamefont
  {Doherty}(2002)}]{MabDoh2002}%
  \BibitemOpen
  \bibfield  {author} {\bibinfo {author} {\bibfnamefont {H.}~\bibnamefont
  {Mabuchi}}\ and\ \bibinfo {author} {\bibfnamefont {A.~C.}\ \bibnamefont
  {Doherty}},\ }\href {\doibase 10.1126/science.1078446} {\bibfield  {journal}
  {\bibinfo  {journal} {Science}\ }\textbf {\bibinfo {volume} {298}},\ \bibinfo
  {pages} {1372} (\bibinfo {year} {2002})}\BibitemShut {NoStop}%
\bibitem [{\citenamefont {Walther}\ \emph {et~al.}(2006)\citenamefont
  {Walther}, \citenamefont {Varcoe}, \citenamefont {Englert},\ and\
  \citenamefont {Becker}}]{Walter2006}%
  \BibitemOpen
  \bibfield  {author} {\bibinfo {author} {\bibfnamefont {H.}~\bibnamefont
  {Walther}}, \bibinfo {author} {\bibfnamefont {B.~T.}\ \bibnamefont {Varcoe}},
  \bibinfo {author} {\bibfnamefont {B.-G.}\ \bibnamefont {Englert}}, \ and\
  \bibinfo {author} {\bibfnamefont {T.}~\bibnamefont {Becker}},\ }\href
  {\doibase 10.1088/0034-4885/69/5/R02} {\bibfield  {journal} {\bibinfo
  {journal} {Rep. Prog. Phys.}\ }\textbf {\bibinfo {volume} {69}},\ \bibinfo
  {pages} {1325} (\bibinfo {year} {2006})}\BibitemShut {NoStop}%
\bibitem [{\citenamefont {Wallraff}\ \emph {et~al.}(2004)\citenamefont
  {Wallraff}, \citenamefont {Schuster}, \citenamefont {Blais}, \citenamefont
  {Frunzio}, \citenamefont {Huang}, \citenamefont {Majer}, \citenamefont
  {Kumar}, \citenamefont {Girvin},\ and\ \citenamefont
  {Schoelkopf}}]{Wallraff2004}%
  \BibitemOpen
  \bibfield  {author} {\bibinfo {author} {\bibfnamefont {A.}~\bibnamefont
  {Wallraff}}, \bibinfo {author} {\bibfnamefont {D.~I.}\ \bibnamefont
  {Schuster}}, \bibinfo {author} {\bibfnamefont {A.}~\bibnamefont {Blais}},
  \bibinfo {author} {\bibfnamefont {L.}~\bibnamefont {Frunzio}}, \bibinfo
  {author} {\bibfnamefont {R.-S.}\ \bibnamefont {Huang}}, \bibinfo {author}
  {\bibfnamefont {J.}~\bibnamefont {Majer}}, \bibinfo {author} {\bibfnamefont
  {S.}~\bibnamefont {Kumar}}, \bibinfo {author} {\bibfnamefont {S.~M.}\
  \bibnamefont {Girvin}}, \ and\ \bibinfo {author} {\bibfnamefont {R.~J.}\
  \bibnamefont {Schoelkopf}},\ }\href {\doibase 10.1038/nature02851} {\bibfield
   {journal} {\bibinfo  {journal} {Nature}\ }\textbf {\bibinfo {volume}
  {431}},\ \bibinfo {pages} {162} (\bibinfo {year} {2004})}\BibitemShut
  {NoStop}%
\bibitem [{\citenamefont {Blais}\ \emph {et~al.}(2004)\citenamefont {Blais},
  \citenamefont {Huang}, \citenamefont {Wallraff}, \citenamefont {Girvin},\
  and\ \citenamefont {Schoelkopf}}]{Blais2004}%
  \BibitemOpen
  \bibfield  {author} {\bibinfo {author} {\bibfnamefont {A.}~\bibnamefont
  {Blais}}, \bibinfo {author} {\bibfnamefont {R.-S.}\ \bibnamefont {Huang}},
  \bibinfo {author} {\bibfnamefont {A.}~\bibnamefont {Wallraff}}, \bibinfo
  {author} {\bibfnamefont {S.~M.}\ \bibnamefont {Girvin}}, \ and\ \bibinfo
  {author} {\bibfnamefont {R.~J.}\ \bibnamefont {Schoelkopf}},\ }\href
  {\doibase 10.1103/PhysRevA.69.062320} {\bibfield  {journal} {\bibinfo
  {journal} {Phys. Rev. A}\ }\textbf {\bibinfo {volume} {69}},\ \bibinfo
  {pages} {062320} (\bibinfo {year} {2004})}\BibitemShut {NoStop}%
\bibitem [{\citenamefont {Frey}\ \emph {et~al.}(2012)\citenamefont {Frey},
  \citenamefont {Leek}, \citenamefont {Beck}, \citenamefont {Blais},
  \citenamefont {Ihn}, \citenamefont {Ensslin},\ and\ \citenamefont
  {Wallraff}}]{Frey2012}%
  \BibitemOpen
  \bibfield  {author} {\bibinfo {author} {\bibfnamefont {T.}~\bibnamefont
  {Frey}}, \bibinfo {author} {\bibfnamefont {P.~J.}\ \bibnamefont {Leek}},
  \bibinfo {author} {\bibfnamefont {M.}~\bibnamefont {Beck}}, \bibinfo {author}
  {\bibfnamefont {A.}~\bibnamefont {Blais}}, \bibinfo {author} {\bibfnamefont
  {T.}~\bibnamefont {Ihn}}, \bibinfo {author} {\bibfnamefont {K.}~\bibnamefont
  {Ensslin}}, \ and\ \bibinfo {author} {\bibfnamefont {A.}~\bibnamefont
  {Wallraff}},\ }\href {\doibase 10.1103/PhysRevLett.108.046807} {\bibfield
  {journal} {\bibinfo  {journal} {Phys. Rev. Lett.}\ }\textbf {\bibinfo
  {volume} {108}},\ \bibinfo {pages} {046807} (\bibinfo {year}
  {2012})}\BibitemShut {NoStop}%
\bibitem [{\citenamefont {Delbecq}\ \emph {et~al.}(2011)\citenamefont
  {Delbecq}, \citenamefont {Schmitt}, \citenamefont {Parmentier}, \citenamefont
  {Roch}, \citenamefont {Viennot}, \citenamefont {F\`eve}, \citenamefont
  {Huard}, \citenamefont {Mora}, \citenamefont {Cottet},\ and\ \citenamefont
  {Kontos}}]{Delbecq2011}%
  \BibitemOpen
  \bibfield  {author} {\bibinfo {author} {\bibfnamefont {M.~R.}\ \bibnamefont
  {Delbecq}}, \bibinfo {author} {\bibfnamefont {V.}~\bibnamefont {Schmitt}},
  \bibinfo {author} {\bibfnamefont {F.~D.}\ \bibnamefont {Parmentier}},
  \bibinfo {author} {\bibfnamefont {N.}~\bibnamefont {Roch}}, \bibinfo {author}
  {\bibfnamefont {J.~J.}\ \bibnamefont {Viennot}}, \bibinfo {author}
  {\bibfnamefont {G.}~\bibnamefont {F\`eve}}, \bibinfo {author} {\bibfnamefont
  {B.}~\bibnamefont {Huard}}, \bibinfo {author} {\bibfnamefont
  {C.}~\bibnamefont {Mora}}, \bibinfo {author} {\bibfnamefont {A.}~\bibnamefont
  {Cottet}}, \ and\ \bibinfo {author} {\bibfnamefont {T.}~\bibnamefont
  {Kontos}},\ }\href {\doibase 10.1103/PhysRevLett.107.256804} {\bibfield
  {journal} {\bibinfo  {journal} {Phys. Rev. Lett.}\ }\textbf {\bibinfo
  {volume} {107}},\ \bibinfo {pages} {256804} (\bibinfo {year}
  {2011})}\BibitemShut {NoStop}%
\bibitem [{\citenamefont {Petersson}\ \emph {et~al.}(2012)\citenamefont
  {Petersson}, \citenamefont {McFaul}, \citenamefont {Schroer}, \citenamefont
  {Jung}, \citenamefont {Taylor}, \citenamefont {Houck},\ and\ \citenamefont
  {Petta}}]{Petersson2012}%
  \BibitemOpen
  \bibfield  {author} {\bibinfo {author} {\bibfnamefont {K.~D.}\ \bibnamefont
  {Petersson}}, \bibinfo {author} {\bibfnamefont {L.~W.}\ \bibnamefont
  {McFaul}}, \bibinfo {author} {\bibfnamefont {M.~D.}\ \bibnamefont {Schroer}},
  \bibinfo {author} {\bibfnamefont {M.}~\bibnamefont {Jung}}, \bibinfo {author}
  {\bibfnamefont {J.~M.}\ \bibnamefont {Taylor}}, \bibinfo {author}
  {\bibfnamefont {A.~A.}\ \bibnamefont {Houck}}, \ and\ \bibinfo {author}
  {\bibfnamefont {J.~R.}\ \bibnamefont {Petta}},\ }\href {\doibase
  10.1038/nature11559} {\bibfield  {journal} {\bibinfo  {journal} {Nature}\
  }\textbf {\bibinfo {volume} {490}},\ \bibinfo {pages} {380} (\bibinfo {year}
  {2012})}\BibitemShut {NoStop}%
\bibitem [{\citenamefont {Liu}\ \emph {et~al.}(2014)\citenamefont {Liu},
  \citenamefont {Petersson}, \citenamefont {Stehlik}, \citenamefont {Taylor},\
  and\ \citenamefont {Petta}}]{Liu2014}%
  \BibitemOpen
  \bibfield  {author} {\bibinfo {author} {\bibfnamefont {Y.-Y.}\ \bibnamefont
  {Liu}}, \bibinfo {author} {\bibfnamefont {K.~D.}\ \bibnamefont {Petersson}},
  \bibinfo {author} {\bibfnamefont {J.}~\bibnamefont {Stehlik}}, \bibinfo
  {author} {\bibfnamefont {J.~M.}\ \bibnamefont {Taylor}}, \ and\ \bibinfo
  {author} {\bibfnamefont {J.~R.}\ \bibnamefont {Petta}},\ }\href {\doibase
  10.1103/PhysRevLett.113.036801} {\bibfield  {journal} {\bibinfo  {journal}
  {Phys. Rev. Lett.}\ }\textbf {\bibinfo {volume} {113}},\ \bibinfo {pages}
  {036801} (\bibinfo {year} {2014})}\BibitemShut {NoStop}%
\bibitem [{\citenamefont {Schwartz}\ \emph {et~al.}(2011)\citenamefont
  {Schwartz}, \citenamefont {Hutchison}, \citenamefont {Genet},\ and\
  \citenamefont {Ebbesen}}]{Schwartz2011}%
  \BibitemOpen
  \bibfield  {author} {\bibinfo {author} {\bibfnamefont {T.}~\bibnamefont
  {Schwartz}}, \bibinfo {author} {\bibfnamefont {J.~A.}\ \bibnamefont
  {Hutchison}}, \bibinfo {author} {\bibfnamefont {C.}~\bibnamefont {Genet}}, \
  and\ \bibinfo {author} {\bibfnamefont {T.~W.}\ \bibnamefont {Ebbesen}},\
  }\href {\doibase 10.1103/PhysRevLett.106.196405} {\bibfield  {journal}
  {\bibinfo  {journal} {Phys. Rev. Lett.}\ }\textbf {\bibinfo {volume} {106}},\
  \bibinfo {pages} {196405} (\bibinfo {year} {2011})}\BibitemShut {NoStop}%
\bibitem [{\citenamefont {Hutchison}\ \emph {et~al.}(2012)\citenamefont
  {Hutchison}, \citenamefont {Schwartz}, \citenamefont {Genet}, \citenamefont
  {Devaux},\ and\ \citenamefont {Ebbesen}}]{Hutchison2012}%
  \BibitemOpen
  \bibfield  {author} {\bibinfo {author} {\bibfnamefont {J.~A.}\ \bibnamefont
  {Hutchison}}, \bibinfo {author} {\bibfnamefont {T.}~\bibnamefont {Schwartz}},
  \bibinfo {author} {\bibfnamefont {C.}~\bibnamefont {Genet}}, \bibinfo
  {author} {\bibfnamefont {E.}~\bibnamefont {Devaux}}, \ and\ \bibinfo {author}
  {\bibfnamefont {T.~W.}\ \bibnamefont {Ebbesen}},\ }\href {\doibase
  10.1002/anie.201107033} {\bibfield  {journal} {\bibinfo  {journal} {Angew.
  Chem. Int. Ed.}\ }\textbf {\bibinfo {volume} {51}},\ \bibinfo {pages} {1592}
  (\bibinfo {year} {2012})}\BibitemShut {NoStop}%
\bibitem [{\citenamefont {Orgiu}\ \emph {et~al.}(2015)\citenamefont {Orgiu},
  \citenamefont {George}, \citenamefont {Hutchison}, \citenamefont {Devaux},
  \citenamefont {Dayen}, \citenamefont {Doudin}, \citenamefont {Stellacci},
  \citenamefont {Genet}, \citenamefont {Schachenmayer}, \citenamefont {Genes},
  \citenamefont {Pupillo}, \citenamefont {Samori},\ and\ \citenamefont
  {Ebbesen}}]{Orgiu2015}%
  \BibitemOpen
  \bibfield  {author} {\bibinfo {author} {\bibfnamefont {E.}~\bibnamefont
  {Orgiu}}, \bibinfo {author} {\bibfnamefont {J.}~\bibnamefont {George}},
  \bibinfo {author} {\bibfnamefont {J.~A.}\ \bibnamefont {Hutchison}}, \bibinfo
  {author} {\bibfnamefont {E.}~\bibnamefont {Devaux}}, \bibinfo {author}
  {\bibfnamefont {J.~F.}\ \bibnamefont {Dayen}}, \bibinfo {author}
  {\bibfnamefont {B.}~\bibnamefont {Doudin}}, \bibinfo {author} {\bibfnamefont
  {F.}~\bibnamefont {Stellacci}}, \bibinfo {author} {\bibfnamefont
  {C.}~\bibnamefont {Genet}}, \bibinfo {author} {\bibfnamefont
  {J.}~\bibnamefont {Schachenmayer}}, \bibinfo {author} {\bibfnamefont
  {C.}~\bibnamefont {Genes}}, \bibinfo {author} {\bibfnamefont
  {G.}~\bibnamefont {Pupillo}}, \bibinfo {author} {\bibfnamefont
  {P.}~\bibnamefont {Samori}}, \ and\ \bibinfo {author} {\bibfnamefont {T.~W.}\
  \bibnamefont {Ebbesen}},\ }\href {http://arxiv.org/abs/1409.1900} {\bibfield
  {journal} {\bibinfo  {journal} {Nature Materials}\ }\textbf {\bibinfo
  {volume} {14}},\ \bibinfo {pages} {1123} (\bibinfo {year} {2015})},\ \bibinfo
  {note} {arXiv:1409.1900}\BibitemShut {NoStop}%
\bibitem [{\citenamefont {Ebbesen}(2016)}]{Ebbesen2016}%
  \BibitemOpen
  \bibfield  {author} {\bibinfo {author} {\bibfnamefont {T.~W.}\ \bibnamefont
  {Ebbesen}},\ }\href {\doibase 10.1021/acs.accounts.6b00295} {\bibfield
  {journal} {\bibinfo  {journal} {Accounts of Chemical Research}\ }\textbf
  {\bibinfo {volume} {49}},\ \bibinfo {pages} {2403} (\bibinfo {year}
  {2016})}\BibitemShut {NoStop}%
\bibitem [{\citenamefont {Zhong}\ \emph {et~al.}(2017)\citenamefont {Zhong},
  \citenamefont {Chervy}, \citenamefont {Zhang}, \citenamefont {Thomas},
  \citenamefont {George}, \citenamefont {Genet}, \citenamefont {Hutchison},\
  and\ \citenamefont {Ebbesen}}]{Zhong2017}%
  \BibitemOpen
  \bibfield  {author} {\bibinfo {author} {\bibfnamefont {X.}~\bibnamefont
  {Zhong}}, \bibinfo {author} {\bibfnamefont {T.}~\bibnamefont {Chervy}},
  \bibinfo {author} {\bibfnamefont {L.}~\bibnamefont {Zhang}}, \bibinfo
  {author} {\bibfnamefont {A.}~\bibnamefont {Thomas}}, \bibinfo {author}
  {\bibfnamefont {J.}~\bibnamefont {George}}, \bibinfo {author} {\bibfnamefont
  {C.}~\bibnamefont {Genet}}, \bibinfo {author} {\bibfnamefont {J.~A.}\
  \bibnamefont {Hutchison}}, \ and\ \bibinfo {author} {\bibfnamefont {T.~W.}\
  \bibnamefont {Ebbesen}},\ }\href {\doibase 10.1002/anie.201703539} {\bibfield
   {journal} {\bibinfo  {journal} {Angewandte Chemie International Edition}\
  }\textbf {\bibinfo {volume} {56}},\ \bibinfo {pages} {9034} (\bibinfo {year}
  {2017})}\BibitemShut {NoStop}%
\bibitem [{\citenamefont {Feist}\ \emph {et~al.}(2018)\citenamefont {Feist},
  \citenamefont {Galego},\ and\ \citenamefont {Garcia-Vidal}}]{FeiGalGar2018}%
  \BibitemOpen
  \bibfield  {author} {\bibinfo {author} {\bibfnamefont {J.}~\bibnamefont
  {Feist}}, \bibinfo {author} {\bibfnamefont {J.}~\bibnamefont {Galego}}, \
  and\ \bibinfo {author} {\bibfnamefont {F.~J.}\ \bibnamefont {Garcia-Vidal}},\
  }\href {\doibase 10.1021/acsphotonics.7b00680} {\bibfield  {journal}
  {\bibinfo  {journal} {ACS Photonics}\ }\textbf {\bibinfo {volume} {5}},\
  \bibinfo {pages} {205} (\bibinfo {year} {2018})}\BibitemShut {NoStop}%
\bibitem [{\citenamefont {Anoop}\ \emph {et~al.}(2016)\citenamefont {Anoop},
  \citenamefont {Jino}, \citenamefont {Atef}, \citenamefont {Marian},
  \citenamefont {J.}, \citenamefont {Joseph}, \citenamefont {Thibault},
  \citenamefont {Xiaolan}, \citenamefont {Elo\"ise}, \citenamefont {Cyriaque},
  \citenamefont {A.},\ and\ \citenamefont {W.}}]{Anoop2016}%
  \BibitemOpen
  \bibfield  {author} {\bibinfo {author} {\bibfnamefont {T.}~\bibnamefont
  {Anoop}}, \bibinfo {author} {\bibfnamefont {G.}~\bibnamefont {Jino}},
  \bibinfo {author} {\bibfnamefont {S.}~\bibnamefont {Atef}}, \bibinfo {author}
  {\bibfnamefont {D.}~\bibnamefont {Marian}}, \bibinfo {author} {\bibfnamefont
  {V.~S.}\ \bibnamefont {J.}}, \bibinfo {author} {\bibfnamefont
  {M.}~\bibnamefont {Joseph}}, \bibinfo {author} {\bibfnamefont
  {C.}~\bibnamefont {Thibault}}, \bibinfo {author} {\bibfnamefont
  {Z.}~\bibnamefont {Xiaolan}}, \bibinfo {author} {\bibfnamefont
  {D.}~\bibnamefont {Elo\"ise}}, \bibinfo {author} {\bibfnamefont
  {G.}~\bibnamefont {Cyriaque}}, \bibinfo {author} {\bibfnamefont {H.~J.}\
  \bibnamefont {A.}}, \ and\ \bibinfo {author} {\bibfnamefont {E.~T.}\
  \bibnamefont {W.}},\ }\href {\doibase 10.1002/anie.201605504} {\bibfield
  {journal} {\bibinfo  {journal} {Angewandte Chemie International Edition}\
  }\textbf {\bibinfo {volume} {55}},\ \bibinfo {pages} {11462} (\bibinfo {year}
  {2016})}\BibitemShut {NoStop}%
\bibitem [{\citenamefont {Tokatly}(2013)}]{Tokatly2013PRL}%
  \BibitemOpen
  \bibfield  {author} {\bibinfo {author} {\bibfnamefont {I.~V.}\ \bibnamefont
  {Tokatly}},\ }\href {\doibase 10.1103/PhysRevLett.110.233001} {\bibfield
  {journal} {\bibinfo  {journal} {Phys. Rev. Lett.}\ }\textbf {\bibinfo
  {volume} {110}},\ \bibinfo {pages} {233001} (\bibinfo {year}
  {2013})}\BibitemShut {NoStop}%
\bibitem [{\citenamefont {Ruggenthaler}\ \emph {et~al.}(2014)\citenamefont
  {Ruggenthaler}, \citenamefont {Flick}, \citenamefont {Pellegrini},
  \citenamefont {Appel}, \citenamefont {Tokatly},\ and\ \citenamefont
  {Rubio}}]{Ruggenthaler2014PRA}%
  \BibitemOpen
  \bibfield  {author} {\bibinfo {author} {\bibfnamefont {M.}~\bibnamefont
  {Ruggenthaler}}, \bibinfo {author} {\bibfnamefont {J.}~\bibnamefont {Flick}},
  \bibinfo {author} {\bibfnamefont {C.}~\bibnamefont {Pellegrini}}, \bibinfo
  {author} {\bibfnamefont {H.}~\bibnamefont {Appel}}, \bibinfo {author}
  {\bibfnamefont {I.~V.}\ \bibnamefont {Tokatly}}, \ and\ \bibinfo {author}
  {\bibfnamefont {A.}~\bibnamefont {Rubio}},\ }\href {\doibase
  10.1103/PhysRevA.90.012508} {\bibfield  {journal} {\bibinfo  {journal} {Phys.
  Rev. A}\ }\textbf {\bibinfo {volume} {90}},\ \bibinfo {pages} {012508}
  (\bibinfo {year} {2014})}\BibitemShut {NoStop}%
\bibitem [{\citenamefont {Flick}\ \emph {et~al.}(2015)\citenamefont {Flick},
  \citenamefont {Ruggenthaler}, \citenamefont {Appel},\ and\ \citenamefont
  {Rubio}}]{Flick2015}%
  \BibitemOpen
  \bibfield  {author} {\bibinfo {author} {\bibfnamefont {J.}~\bibnamefont
  {Flick}}, \bibinfo {author} {\bibfnamefont {M.}~\bibnamefont {Ruggenthaler}},
  \bibinfo {author} {\bibfnamefont {H.}~\bibnamefont {Appel}}, \ and\ \bibinfo
  {author} {\bibfnamefont {A.}~\bibnamefont {Rubio}},\ }\href {\doibase
  10.1073/pnas.1518224112} {\bibfield  {journal} {\bibinfo  {journal} {PNAS}\
  }\textbf {\bibinfo {volume} {112}},\ \bibinfo {pages} {15285} (\bibinfo
  {year} {2015})}\BibitemShut {NoStop}%
\bibitem [{\citenamefont {Galego}\ \emph {et~al.}(2015)\citenamefont {Galego},
  \citenamefont {Garcia-Vidal},\ and\ \citenamefont {Feist}}]{GalGarFei2015}%
  \BibitemOpen
  \bibfield  {author} {\bibinfo {author} {\bibfnamefont {J.}~\bibnamefont
  {Galego}}, \bibinfo {author} {\bibfnamefont {F.~J.}\ \bibnamefont
  {Garcia-Vidal}}, \ and\ \bibinfo {author} {\bibfnamefont {J.}~\bibnamefont
  {Feist}},\ }\href {\doibase 10.1103/PhysRevX.5.041022} {\bibfield  {journal}
  {\bibinfo  {journal} {Phys. Rev. X}\ }\textbf {\bibinfo {volume} {5}},\
  \bibinfo {pages} {041022} (\bibinfo {year} {2015})}\BibitemShut {NoStop}%
\bibitem [{\citenamefont {Pellegrini}\ \emph {et~al.}(2015)\citenamefont
  {Pellegrini}, \citenamefont {Flick}, \citenamefont {Tokatly}, \citenamefont
  {Appel},\ and\ \citenamefont {Rubio}}]{Pellegrini2015PRL}%
  \BibitemOpen
  \bibfield  {author} {\bibinfo {author} {\bibfnamefont {C.}~\bibnamefont
  {Pellegrini}}, \bibinfo {author} {\bibfnamefont {J.}~\bibnamefont {Flick}},
  \bibinfo {author} {\bibfnamefont {I.~V.}\ \bibnamefont {Tokatly}}, \bibinfo
  {author} {\bibfnamefont {H.}~\bibnamefont {Appel}}, \ and\ \bibinfo {author}
  {\bibfnamefont {A.}~\bibnamefont {Rubio}},\ }\href {\doibase
  10.1103/PhysRevLett.115.093001} {\bibfield  {journal} {\bibinfo  {journal}
  {Phys. Rev. Lett.}\ }\textbf {\bibinfo {volume} {115}},\ \bibinfo {pages}
  {093001} (\bibinfo {year} {2015})}\BibitemShut {NoStop}%
\bibitem [{\citenamefont {Kowalewski}\ \emph {et~al.}(2016)\citenamefont
  {Kowalewski}, \citenamefont {Bennett},\ and\ \citenamefont
  {Mukamel}}]{KowBenMuk2016}%
  \BibitemOpen
  \bibfield  {author} {\bibinfo {author} {\bibfnamefont {M.}~\bibnamefont
  {Kowalewski}}, \bibinfo {author} {\bibfnamefont {K.}~\bibnamefont {Bennett}},
  \ and\ \bibinfo {author} {\bibfnamefont {S.}~\bibnamefont {Mukamel}},\ }\href
  {\doibase 10.1021/acs.jpclett.6b00864} {\bibfield  {journal} {\bibinfo
  {journal} {The Journal of Physical Chemistry Letters}\ }\textbf {\bibinfo
  {volume} {7}},\ \bibinfo {pages} {2050} (\bibinfo {year} {2016})},\ \bibinfo
  {note} {pMID: 27186666}\BibitemShut {NoStop}%
\bibitem [{\citenamefont {Galego}\ \emph {et~al.}(2016)\citenamefont {Galego},
  \citenamefont {Garcia-Vidal},\ and\ \citenamefont {Feist}}]{GalGarFei2016}%
  \BibitemOpen
  \bibfield  {author} {\bibinfo {author} {\bibfnamefont {J.}~\bibnamefont
  {Galego}}, \bibinfo {author} {\bibfnamefont {F.~J.}\ \bibnamefont
  {Garcia-Vidal}}, \ and\ \bibinfo {author} {\bibfnamefont {J.}~\bibnamefont
  {Feist}},\ }\href {\doibase 10.1038/ncomms13841} {\bibfield  {journal}
  {\bibinfo  {journal} {Nature Comm.}\ }\textbf {\bibinfo {volume} {7}},\
  \bibinfo {pages} {13841} (\bibinfo {year} {2016})}\BibitemShut {NoStop}%
\bibitem [{\citenamefont {Galego}\ \emph {et~al.}(2017)\citenamefont {Galego},
  \citenamefont {Garcia-Vidal},\ and\ \citenamefont {Feist}}]{GalGarFei2017}%
  \BibitemOpen
  \bibfield  {author} {\bibinfo {author} {\bibfnamefont {J.}~\bibnamefont
  {Galego}}, \bibinfo {author} {\bibfnamefont {F.~J.}\ \bibnamefont
  {Garcia-Vidal}}, \ and\ \bibinfo {author} {\bibfnamefont {J.}~\bibnamefont
  {Feist}},\ }\href {\doibase 10.1103/PhysRevLett.119.136001} {\bibfield
  {journal} {\bibinfo  {journal} {Phys. Rev. Lett.}\ }\textbf {\bibinfo
  {volume} {119}},\ \bibinfo {pages} {136001} (\bibinfo {year}
  {2017})}\BibitemShut {NoStop}%
\bibitem [{\citenamefont {Flick}\ \emph
  {et~al.}(2017{\natexlab{a}})\citenamefont {Flick}, \citenamefont
  {Ruggenthaler}, \citenamefont {Appel},\ and\ \citenamefont
  {Rubio}}]{Flick2017a}%
  \BibitemOpen
  \bibfield  {author} {\bibinfo {author} {\bibfnamefont {J.}~\bibnamefont
  {Flick}}, \bibinfo {author} {\bibfnamefont {M.}~\bibnamefont {Ruggenthaler}},
  \bibinfo {author} {\bibfnamefont {H.}~\bibnamefont {Appel}}, \ and\ \bibinfo
  {author} {\bibfnamefont {A.}~\bibnamefont {Rubio}},\ }\href {\doibase
  10.1073/pnas.1615509114} {\bibfield  {journal} {\bibinfo  {journal} {PNAS}\
  }\textbf {\bibinfo {volume} {114}},\ \bibinfo {pages} {3026} (\bibinfo {year}
  {2017}{\natexlab{a}})}\BibitemShut {NoStop}%
\bibitem [{\citenamefont {Flick}\ \emph
  {et~al.}(2017{\natexlab{b}})\citenamefont {Flick}, \citenamefont {Appel},
  \citenamefont {Ruggenthaler},\ and\ \citenamefont {Rubio}}]{Flick2017b}%
  \BibitemOpen
  \bibfield  {author} {\bibinfo {author} {\bibfnamefont {J.}~\bibnamefont
  {Flick}}, \bibinfo {author} {\bibfnamefont {H.}~\bibnamefont {Appel}},
  \bibinfo {author} {\bibfnamefont {M.}~\bibnamefont {Ruggenthaler}}, \ and\
  \bibinfo {author} {\bibfnamefont {A.}~\bibnamefont {Rubio}},\ }\href
  {\doibase 10.1021/acs.jctc.6b01126} {\bibfield  {journal} {\bibinfo
  {journal} {J. Chem. Theory Comput.}\ }\textbf {\bibinfo {volume} {13}},\
  \bibinfo {pages} {1616} (\bibinfo {year} {2017}{\natexlab{b}})},\ \bibinfo
  {note} {pMID: 28277664}\BibitemShut {NoStop}%
\bibitem [{\citenamefont {Flick}\ \emph
  {et~al.}(2018{\natexlab{a}})\citenamefont {Flick}, \citenamefont {Sch\"afer},
  \citenamefont {Ruggenthaler}, \citenamefont {Appel},\ and\ \citenamefont
  {Rubio}}]{Flick2018}%
  \BibitemOpen
  \bibfield  {author} {\bibinfo {author} {\bibfnamefont {J.}~\bibnamefont
  {Flick}}, \bibinfo {author} {\bibfnamefont {C.}~\bibnamefont {Sch\"afer}},
  \bibinfo {author} {\bibfnamefont {M.}~\bibnamefont {Ruggenthaler}}, \bibinfo
  {author} {\bibfnamefont {H.}~\bibnamefont {Appel}}, \ and\ \bibinfo {author}
  {\bibfnamefont {A.}~\bibnamefont {Rubio}},\ }\href {\doibase
  10.1021/acsphotonics.7b01279} {\bibfield  {journal} {\bibinfo  {journal} {ACS
  Photonics}\ }\textbf {\bibinfo {volume} {5}},\ \bibinfo {pages} {992}
  (\bibinfo {year} {2018}{\natexlab{a}})}\BibitemShut {NoStop}%
\bibitem [{\citenamefont {Vendrell}(2018)}]{Vendrell2018}%
  \BibitemOpen
  \bibfield  {author} {\bibinfo {author} {\bibfnamefont {O.}~\bibnamefont
  {Vendrell}},\ }\href {\doibase
  https://doi.org/10.1016/j.chemphys.2018.02.008} {\bibfield  {journal}
  {\bibinfo  {journal} {Chemical Physics}\ }\textbf {\bibinfo {volume} {509}},\
  \bibinfo {pages} {55 } (\bibinfo {year} {2018})},\ \bibinfo {note}
  {high-dimensional quantum dynamics (on the occasion of the 70th birthday of
  Hans-Dieter Meyer)}\BibitemShut {NoStop}%
\bibitem [{\citenamefont {Flick}\ and\ \citenamefont
  {Narang}(2018)}]{FliNar2018}%
  \BibitemOpen
  \bibfield  {author} {\bibinfo {author} {\bibfnamefont {J.}~\bibnamefont
  {Flick}}\ and\ \bibinfo {author} {\bibfnamefont {P.}~\bibnamefont {Narang}},\
  }\href {\doibase 10.1103/PhysRevLett.121.113002} {\bibfield  {journal}
  {\bibinfo  {journal} {Phys. Rev. Lett.}\ }\textbf {\bibinfo {volume} {121}},\
  \bibinfo {pages} {113002} (\bibinfo {year} {2018})}\BibitemShut {NoStop}%
\bibitem [{\citenamefont {Flick}\ \emph
  {et~al.}(2018{\natexlab{b}})\citenamefont {Flick}, \citenamefont {Rivera},\
  and\ \citenamefont {Narang}}]{PinRivNar2018}%
  \BibitemOpen
  \bibfield  {author} {\bibinfo {author} {\bibfnamefont {J.}~\bibnamefont
  {Flick}}, \bibinfo {author} {\bibfnamefont {N.}~\bibnamefont {Rivera}}, \
  and\ \bibinfo {author} {\bibfnamefont {P.}~\bibnamefont {Narang}},\ }\href
  {\doibase 10.1515/nanoph-2018-0067} {\bibfield  {journal} {\bibinfo
  {journal} {Nanophotonics}\ }\textbf {\bibinfo {volume} {7}},\ \bibinfo
  {pages} {1479} (\bibinfo {year} {2018}{\natexlab{b}})}\BibitemShut {NoStop}%
\bibitem [{\citenamefont {Onida}\ \emph {et~al.}(2002)\citenamefont {Onida},
  \citenamefont {Reining},\ and\ \citenamefont {Rubio}}]{OniReiRub2002}%
  \BibitemOpen
  \bibfield  {author} {\bibinfo {author} {\bibfnamefont {G.}~\bibnamefont
  {Onida}}, \bibinfo {author} {\bibfnamefont {L.}~\bibnamefont {Reining}}, \
  and\ \bibinfo {author} {\bibfnamefont {A.}~\bibnamefont {Rubio}},\ }\href
  {\doibase 10.1103/RevModPhys.74.601} {\bibfield  {journal} {\bibinfo
  {journal} {Rev. Mod. Phys.}\ }\textbf {\bibinfo {volume} {74}},\ \bibinfo
  {pages} {601} (\bibinfo {year} {2002})}\BibitemShut {NoStop}%
\bibitem [{\citenamefont {Stefanucci}\ and\ \citenamefont {{van
  Leeuwen}}(2013)}]{Stefanucci-book}%
  \BibitemOpen
  \bibfield  {author} {\bibinfo {author} {\bibfnamefont {G.}~\bibnamefont
  {Stefanucci}}\ and\ \bibinfo {author} {\bibfnamefont {R.}~\bibnamefont {{van
  Leeuwen}}},\ }\href@noop {} {\emph {\bibinfo {title} {Nonequilibrium
  many-body theory of quantum systems}}}\ (\bibinfo  {publisher} {Cambridge
  Univ. Press},\ \bibinfo {address} {Cambridge},\ \bibinfo {year}
  {2013})\BibitemShut {NoStop}%
\bibitem [{\citenamefont {Runge}\ and\ \citenamefont
  {Gross}(1984)}]{RunGro1984}%
  \BibitemOpen
  \bibfield  {author} {\bibinfo {author} {\bibfnamefont {E.}~\bibnamefont
  {Runge}}\ and\ \bibinfo {author} {\bibfnamefont {E.~K.~U.}\ \bibnamefont
  {Gross}},\ }\href {\doibase 10.1103/PhysRevLett.52.997} {\bibfield  {journal}
  {\bibinfo  {journal} {Phys. Rev. Lett.}\ }\textbf {\bibinfo {volume} {52}},\
  \bibinfo {pages} {997} (\bibinfo {year} {1984})}\BibitemShut {NoStop}%
\bibitem [{\citenamefont {Marques}\ \emph {et~al.}(2012)\citenamefont
  {Marques}, \citenamefont {Maitra}, \citenamefont {Nogueira}, \citenamefont
  {Gross},\ and\ \citenamefont {Rubio}}]{TDDFT-2012}%
  \BibitemOpen
  \bibinfo {editor} {\bibfnamefont {M.~A.}\ \bibnamefont {Marques}}, \bibinfo
  {editor} {\bibfnamefont {N.~T.}\ \bibnamefont {Maitra}}, \bibinfo {editor}
  {\bibfnamefont {F.~M.}\ \bibnamefont {Nogueira}}, \bibinfo {editor}
  {\bibfnamefont {E.}~\bibnamefont {Gross}}, \ and\ \bibinfo {editor}
  {\bibfnamefont {A.}~\bibnamefont {Rubio}},\ eds.,\ \href@noop {} {\emph
  {\bibinfo {title} {Fundamentals of Time-Dependent Density Functional
  Theory}}}\ (\bibinfo  {publisher} {Springer},\ \bibinfo {address} {Berlin},\
  \bibinfo {year} {2012})\BibitemShut {NoStop}%
\bibitem [{\citenamefont {Ullrich}(2012)}]{TDDFTbyUllrich}%
  \BibitemOpen
  \bibfield  {author} {\bibinfo {author} {\bibfnamefont {C.~A.}\ \bibnamefont
  {Ullrich}},\ }\href@noop {} {\emph {\bibinfo {title} {Time-Dependent
  Density-Functional Theory: Concepts and Applications}}}\ (\bibinfo
  {publisher} {Oxford University Press},\ \bibinfo {address} {New York},\
  \bibinfo {year} {2012})\BibitemShut {NoStop}%
\bibitem [{\citenamefont {Trevisanutto}\ and\ \citenamefont
  {Milletar\`{\i}}(2015)}]{TreMil2015}%
  \BibitemOpen
  \bibfield  {author} {\bibinfo {author} {\bibfnamefont {P.~E.}\ \bibnamefont
  {Trevisanutto}}\ and\ \bibinfo {author} {\bibfnamefont {M.}~\bibnamefont
  {Milletar\`{\i}}},\ }\href {\doibase 10.1103/PhysRevB.92.235303} {\bibfield
  {journal} {\bibinfo  {journal} {Phys. Rev. B}\ }\textbf {\bibinfo {volume}
  {92}},\ \bibinfo {pages} {235303} (\bibinfo {year} {2015})}\BibitemShut
  {NoStop}%
\bibitem [{\citenamefont {de~Melo}\ and\ \citenamefont
  {Marini}(2016)}]{MelMar2016}%
  \BibitemOpen
  \bibfield  {author} {\bibinfo {author} {\bibfnamefont {P.~M. M.~C.}\
  \bibnamefont {de~Melo}}\ and\ \bibinfo {author} {\bibfnamefont
  {A.}~\bibnamefont {Marini}},\ }\href {\doibase 10.1103/PhysRevB.93.155102}
  {\bibfield  {journal} {\bibinfo  {journal} {Phys. Rev. B}\ }\textbf {\bibinfo
  {volume} {93}},\ \bibinfo {pages} {155102} (\bibinfo {year}
  {2016})}\BibitemShut {NoStop}%
\bibitem [{\citenamefont {Baym}(1962)}]{Baym1962}%
  \BibitemOpen
  \bibfield  {author} {\bibinfo {author} {\bibfnamefont {G.}~\bibnamefont
  {Baym}},\ }\href@noop {} {\bibfield  {journal} {\bibinfo  {journal} {Phys.
  Rev.}\ }\textbf {\bibinfo {volume} {127}},\ \bibinfo {pages} {1391 }
  (\bibinfo {year} {1962})}\BibitemShut {NoStop}%
\bibitem [{\citenamefont {Dobson}(1994)}]{Dobson1994}%
  \BibitemOpen
  \bibfield  {author} {\bibinfo {author} {\bibfnamefont {J.~F.}\ \bibnamefont
  {Dobson}},\ }\href@noop {} {\bibfield  {journal} {\bibinfo  {journal} {Phys.
  Rev. Lett.}\ }\textbf {\bibinfo {volume} {73}},\ \bibinfo {pages} {2244}
  (\bibinfo {year} {1994})}\BibitemShut {NoStop}%
\bibitem [{\citenamefont {Vignale}(1995{\natexlab{a}})}]{Vignale1995a}%
  \BibitemOpen
  \bibfield  {author} {\bibinfo {author} {\bibfnamefont {G.}~\bibnamefont
  {Vignale}},\ }\href@noop {} {\bibfield  {journal} {\bibinfo  {journal} {Phys.
  Rev. Lett.}\ }\textbf {\bibinfo {volume} {74}},\ \bibinfo {pages} {3233}
  (\bibinfo {year} {1995}{\natexlab{a}})}\BibitemShut {NoStop}%
\bibitem [{\citenamefont {Vignale}(1995{\natexlab{b}})}]{Vignale1995b}%
  \BibitemOpen
  \bibfield  {author} {\bibinfo {author} {\bibfnamefont {G.}~\bibnamefont
  {Vignale}},\ }\href@noop {} {\bibfield  {journal} {\bibinfo  {journal} {Phys.
  Lett. A}\ }\textbf {\bibinfo {volume} {209}},\ \bibinfo {pages} {206}
  (\bibinfo {year} {1995}{\natexlab{b}})}\BibitemShut {NoStop}%
\bibitem [{\citenamefont {von Barth}\ \emph {et~al.}(2005)\citenamefont {von
  Barth}, \citenamefont {Dahlen}, \citenamefont {van Leeuwen},\ and\
  \citenamefont {Stefanucci}}]{Barth2005}%
  \BibitemOpen
  \bibfield  {author} {\bibinfo {author} {\bibfnamefont {U.}~\bibnamefont {von
  Barth}}, \bibinfo {author} {\bibfnamefont {N.~E.}\ \bibnamefont {Dahlen}},
  \bibinfo {author} {\bibfnamefont {R.}~\bibnamefont {van Leeuwen}}, \ and\
  \bibinfo {author} {\bibfnamefont {G.}~\bibnamefont {Stefanucci}},\ }\href
  {\doibase 10.1103/PhysRevB.72.235109} {\bibfield  {journal} {\bibinfo
  {journal} {Phys. Rev. B}\ }\textbf {\bibinfo {volume} {72}},\ \bibinfo
  {pages} {235109} (\bibinfo {year} {2005})}\BibitemShut {NoStop}%
\bibitem [{\citenamefont {Power}\ and\ \citenamefont
  {Zienau}(1959)}]{PowZie1959}%
  \BibitemOpen
  \bibfield  {author} {\bibinfo {author} {\bibfnamefont {E.~A.}\ \bibnamefont
  {Power}}\ and\ \bibinfo {author} {\bibfnamefont {S.}~\bibnamefont {Zienau}},\
  }\href {\doibase 10.1098/rsta.1959.0008} {\bibfield  {journal} {\bibinfo
  {journal} {Philosophical Transactions of the Royal Society of London A:
  Mathematical, Physical and Engineering Sciences}\ }\textbf {\bibinfo {volume}
  {251}},\ \bibinfo {pages} {427} (\bibinfo {year} {1959})}\BibitemShut
  {NoStop}%
\bibitem [{\citenamefont {Woolley}(1971)}]{Woolley1971}%
  \BibitemOpen
  \bibfield  {author} {\bibinfo {author} {\bibfnamefont {R.~G.}\ \bibnamefont
  {Woolley}},\ }\href {\doibase 10.1098/rspa.1971.0049} {\bibfield  {journal}
  {\bibinfo  {journal} {Proceedings of the Royal Society of London A:
  Mathematical, Physical and Engineering Sciences}\ }\textbf {\bibinfo {volume}
  {321}},\ \bibinfo {pages} {557} (\bibinfo {year} {1971})}\BibitemShut
  {NoStop}%
\bibitem [{\citenamefont {Abedi}\ \emph {et~al.}(2018)\citenamefont {Abedi},
  \citenamefont {Khosravi},\ and\ \citenamefont {Tokatly}}]{AbeKhoTok2018EPJB}%
  \BibitemOpen
  \bibfield  {author} {\bibinfo {author} {\bibfnamefont {A.}~\bibnamefont
  {Abedi}}, \bibinfo {author} {\bibfnamefont {E.}~\bibnamefont {Khosravi}}, \
  and\ \bibinfo {author} {\bibfnamefont {I.~V.}\ \bibnamefont {Tokatly}},\
  }\href {\doibase 10.1140/epjb/e2018-90243-1} {\bibfield  {journal} {\bibinfo
  {journal} {Eur. Phys. J. B}\ }\textbf {\bibinfo {volume} {91}},\ \bibinfo
  {pages} {194} (\bibinfo {year} {2018})}\BibitemShut {NoStop}%
\bibitem [{\citenamefont {Faisal}(1987)}]{MultiphotonbyFaisal}%
  \BibitemOpen
  \bibfield  {author} {\bibinfo {author} {\bibfnamefont {F.~H.~M.}\
  \bibnamefont {Faisal}},\ }\href@noop {} {\emph {\bibinfo {title} {Theory of
  Multiphoton Processes}}}\ (\bibinfo  {publisher} {Plenum Press},\ \bibinfo
  {address} {New York},\ \bibinfo {year} {1987})\BibitemShut {NoStop}%
\bibitem [{\citenamefont {Tokatly}(2005)}]{Tokatly2005PRBa}%
  \BibitemOpen
  \bibfield  {author} {\bibinfo {author} {\bibfnamefont {I.~V.}\ \bibnamefont
  {Tokatly}},\ }\href@noop {} {\bibfield  {journal} {\bibinfo  {journal} {Phys.
  Rev. B}\ }\textbf {\bibinfo {volume} {71}},\ \bibinfo {pages} {165104}
  (\bibinfo {year} {2005})}\BibitemShut {NoStop}%
\bibitem [{\citenamefont {van Leeuwen}\ and\ \citenamefont
  {Stefanucci}(2013)}]{LeeSte2013}%
  \BibitemOpen
  \bibfield  {author} {\bibinfo {author} {\bibfnamefont {R.}~\bibnamefont {van
  Leeuwen}}\ and\ \bibinfo {author} {\bibfnamefont {G.}~\bibnamefont
  {Stefanucci}},\ }\href {http://stacks.iop.org/1742-6596/427/i=1/a=012001}
  {\bibfield  {journal} {\bibinfo  {journal} {Journal of Physics: Conference
  Series}\ }\textbf {\bibinfo {volume} {427}},\ \bibinfo {pages} {012001}
  (\bibinfo {year} {2013})}\BibitemShut {NoStop}%
\bibitem [{\citenamefont {Almbladh}\ \emph {et~al.}(1999)\citenamefont
  {Almbladh}, \citenamefont {von Barth},\ and\ \citenamefont {van
  Leeuwen}}]{AlmBarLee1999}%
  \BibitemOpen
  \bibfield  {author} {\bibinfo {author} {\bibfnamefont {C.-O.}\ \bibnamefont
  {Almbladh}}, \bibinfo {author} {\bibfnamefont {U.}~\bibnamefont {von Barth}},
  \ and\ \bibinfo {author} {\bibfnamefont {R.}~\bibnamefont {van Leeuwen}},\
  }\href@noop {} {\bibfield  {journal} {\bibinfo  {journal} {Int. J. Mod. Phys.
  B}\ }\textbf {\bibinfo {volume} {13}},\ \bibinfo {pages} {535} (\bibinfo
  {year} {1999})}\BibitemShut {NoStop}%
\end{thebibliography}

%

\end{document}